\documentclass[11pt, twocolumn]{aastex631}
\usepackage{CJK}
\usepackage{bm}
\usepackage{units}
\usepackage{amsmath}
\usepackage{float}
\usepackage{pgfplots}
\usepackage{multirow}
\pgfplotsset{compat=1.15}

\date{\today}

\begin{document}
\begin{CJK*}{UTF8}{gbsn}

\title{Interplanetary magnetic correlation and low-frequency spectrum over many solar rotations}
\author[0009-0008-8723-610X]{Jiaming Wang (王嘉明)}
\affiliation{Department of Physics and Astronomy, University of Delaware}
\author[0000-0003-4168-590X]{Francesco Pecora}
\affiliation{Department of Physics and Astronomy, University of Delaware}
\author[0000-0002-7174-6948]{Rohit Chhiber}
\affiliation{Department of Physics and Astronomy, University of Delaware}
\affiliation{Heliophysics Science Division, NASA Goddard Space Flight Center}
\author[0000-0003-3891-5495]{Sohom Roy}
\affiliation{Department of Physics and Astronomy, University of Delaware}
\author[0000-0001-7224-6024]{William H. Matthaeus}
\affiliation{Department of Physics and Astronomy, University of Delaware}


\begin{abstract}
Fluctuations and structure across a wide range of spatial and temporal scales are frequently studied in the solar wind. The properties of the low-frequency fluctuations are of relevance to turbulent energy injection into the plasma and the transport of high-energy cosmic rays. Correlation analysis of decade-long intervals of interplanetary data permits study of fluctuations at time scales much longer than suitably defined correlation times, and therefore at frequencies well below those associated with the Kolmogorov inertial range of {\it in situ} turbulence. At the frequencies of interest, we study the familiar occurrence of the $1/f$ spectral signature. We also study point spectral features due to solar rotation and their relation with the $1/f$ signal. We report novel properties at timescales ranging from minutes up to years, using data selected by wind speed, phase of solar cycle, and cartesian components of the magnetic field. A surprising finding is that the power in solar rotation harmonics is consistent with an extension of the $1/f$ spectrum, down to frequencies as low as around $\unit[5 \times 10^{-7}]{Hz}$. The presence of a broadband $1/f$ spectrum across different wind types supports the interpretation that $1/f$ signals may be related to or even originate from the solar dynamo.
\end{abstract}

\section{Introduction}
\label{sec:intro}
tatistical methods have long been employed to study variability of heliospheric plasma and magnetic fields. Correlation and spectral analysis have been prominent in this regard, particularly in characterizing properties of turbulence~\citep{Bruno13}. Aided by the Taylor hypothesis that approximately relates temporal fluctuations at a fixed point to spatial fluctuations advected by the flow~\citep{Taylor38}, spectral analysis enables study of turbulence 
properties. Prominent among these 
is the familiar 
power-law inertial range
typically interpreted as evidence of a local Kolmogorov-like energy cascade~\citep{Coleman68}.
At lower frequencies, below the measured correlation timescale,
one might expect a flat spectrum indicative of completely uncorrelated signals. Yet, this is not often seen in the interplanetary magnetic field. Instead,
for very low frequencies (typically below $\unit[10^{-3}]{Hz}$), a different feature called $1/f$ noise emerges, 
suggestive of persistent correlation.
Also known as ``flicker noise'', the $1/f$ spectrum has the unique property of distributing equal power across logarithmic frequency intervals, i.e., constant power per octave. 
While 
there exist generic pathways 
to generating the $1/f$
signal in a variety of physical systems~\citep{Machlup81,Dutta81,Montroll82,Bak87}, its origin in the heliospheric plasma remains
a topic of debate (see, e.g., \cite{Matthaeus86, Velli89, Verdini12, Matteini18, Chandran18, Magyar22} for various proposed generation mechanisms; and \cite{Davis23, Huang23, Pradata25} for some recent observations). See \cite{Wang24_1overf} for a recent review.

Some major questions that arise concerning 
the observed very-low frequency interplanetary
fluctuations are:
(i) can $1/f$ noise be generated locally in the interplanetary medium, either by expansion or by turbulence?
(ii) alternatively, does $1/f$ represent effects of  
lower coronal or sub-photospheric dynamical processes that migrate into the solar wind?
(iii) what is the relationship between $1/f$ and 
signals at the solar rotation period and its harmonics?
(iv) do the signals related to 
solar rotation 
represent intrinsic features of the heliospheric system or are they consequences of the position of observation? 
We will address these questions below. 
We also discuss whether $1/f$ observations
might originate in more than one process that act at different frequencies. Finally, we comment on the impact that the phenomena studied here might have on the geospace environment and prediction of 
space weather.

{\it Background on $1/f$ spectral observations in the inner heliosphere.} A feature of the $1/f$ signal that invariably enters the discussion is the range over which it is observed. On  the low-frequency end, it has been  seen to extend down to $\sim 2\times \unit[10^{-6}]{Hz}$~\citep{Matthaeus86}. This frequency is so low that finite propagation speed of local magnetohydrodynamic (MHD) processes make it unlikely that {\it in situ} mechanisms alone can account for the full extent~\citep[][]{Zhou90, Chhiber2018thesis}. However, ideas concerning local effects and wind expansion have been presented to explain the higher frequency part of the $1/f$ spectrum~\citep{Velli89, Huang23}. An early paper examining spectra over a wide frequency range from $\unit[2.4 \times 10^{-5}]{Hz}$
to $\unit[470]{Hz}$, including range of $1/f$, was presented by 
\cite{Denskat83}. For a recent review paper including $1/f$ in other physical systems, see \cite{Wang24_1overf}.

It should be remarked that the 
transition between the $1/f$ range and the familiar $f^{-5/3}$ turbulence inertial range is generally associated with sweeping of 
the correlation scale past the 
detector~\citep{Tu95}. This ``break point'' is typically not sharp but gradual. It is also well
documented to evolve toward lower frequency with increasing heliocentric distance~\citep{Klein92, Tu95, Cuesta22, Davis23}. The break occurs near the correlation scale and
may thus be reasonably interpreted as the low-frequency end of the inertial range. The high frequency edge of the inertial range, 
marking transition to kinetic effects, 
is associated with sweeping of either the  
thermal ion gyroradius or the ion inertial scale~\citep[][]{Chen14}. 
On a theoretical level, 
the temporal development of turbulence 
leads naturally
to production of 
correlations
at progressively larger scales~\citep{deKarman38, Hossain95, Wan12}. 
Therefore, a plausible 
picture is that local Kolmogorov-type turbulence 
in the solar wind 
progressively generates correlated fluctuations, causing the break point between $1/f$ and the inertial range to move toward lower frequency with increasing heliocentric distance. In addition, wind expansion could also contribute to growth of correlations over time in the plane transverse to the radial direction.

In the present paper,
we examine the context of the $1/f$
signal, emphasizing its relationship to the turbulence inertial range and also to signals at very low frequencies, in particular those having
periods related to solar rotation.
We show in Section~\ref{sec:results} the presence of a secondary 13.5-day periodicity superposed on the dominant 27-day solar rotation periodicity, which becomes particularly prominent during the solar minimum phase of the analyzed 23rd solar cycle. Periodic signals on day-long timescales in the solar wind, especially those associated with solar rotation or linked to chromospheric structures, have been extensively studied for their role in shaping solar-terrestrial interactions~\citep[see, e.g., ][]{Xie17}.

{\it Background on solar rotation frequencies.}
An early and rather comprehensive study of low-frequency 
periodicities was carried out in \cite{Gosling76} using 
3-hour averaged solar wind speed data beyond Earth's bow shock from Vela and IMP satellites.
Yearly autocorrelation analysis of solar wind speed reveals a dominant 27.1-day periodicity during 1971, 1973, and 1974, with a 13.5-day periodicity emerging toward the 1974 solar minimum. Concurrently, high-speed ($\geq \unit[650]{km/s}$) streams that are predominantly unipolar (not bounded by magnetic sectors) appear with increasing amplitude and duration toward solar minimum and the declining phase. The observed recurrence of structures is associated with solar cycle phases, particularly the distribution of coronal holes conjectured as the sources of fast solar wind. \cite{Gosling76} attribute the 13.5-day periodicity to two broad and stable high-speed streams roughly $180^\circ$ apart in solar longitude. A similar conclusion is obtained by \cite{Donnelly90}, which find 13- to 14-day periodicities in multi-wavelength chromospheric flux that likely arise from active regions separated by around $180^\circ$ in longitude.


\cite{Fenimore78} study solar wind speed power spectra over a similar solar cycle period as \cite{Gosling76}, and identify not only several harmonics of the 27-day periodicity but also a broad band of spectral peaks spanning 25-31.5 days and 13-14.5 days. While a spread in periodicities could, in principle, arise from differential solar rotation, this mechanism is insufficient to account for the width of the observed bands. Instead, \cite{Fenimore78} attribute the range of spectral peaks to the existence of a set of recurrent high-speed streams with the same frequency but slight phase shifts, corresponding to structures distributed across a band of solar longitudes.

Further observations of the $\sim 13.5$-day periodicity are reported in geomagnetic indices~\citep{Gonzalez93, Syiemlieh22, Nowak24}; solar wind parameters such as speed, temperature, ion density, and interplanetary magnetic field~\citep[][]{Mursula96, Syiemlieh22}; chromospheric spectroscopic measurements~\citep{Mursula96, Kotze23}; and the solar mean magnetic field~\citep[magnetic field strength averaged over the entire solar disk;][]{Xie17}. During solar minimum and the declining phase, the 13.5-day signal can at times exhibit greater power than the 27-day component. Here, we do not seek to explain the origin of these 13.5-day structures, but instead undertake a phenomenological approach to characterize their appearances in the slow and fast solar wind, as well as during solar minimum and maximum. In particular, we focus on the spectral density underlying the 27- and 13.5-day structures as well as the higher-order harmonics and their relationship with the interplanetary $1/f$ spectrum, conventionally observed at higher frequencies~\citep{Matthaeus86}. This is in part motivated by the recent increase in interest in the nature and origin of the heliospheric $1/f$ noise. 

Here we take a fresh look 
at these low-frequency signals in the interplanetary 
magnetic field and the density at 1 au. This is motivated by
(1) the incomplete consensus regarding the origin of the 27-day signal and its harmonics, and also by 
(2) the recent increase in interest in the 
nature and origin of the interplanetary 
$1/f$ signal. We also examine a novel relationship between these two often studied 
phenomena. 

This paper is organized as follows: in Section~\ref{sec:data}, we describe the data employed, the correlation analysis, and the power-spectral analysis. In Section~\ref{sec:results}, we compare results from correlation and spectral analysis over different solar wind categories, including the slow and fast solar wind, as well as during solar maximum and minimum. In Section~\ref{sec:conclusion}, we discuss our results in the context of the discussion posed above.

\section{Data and Analysis Procedure}
\label{sec:data}

We employ 12 years of data (February 1998 to December 2009; around solar minimum to minimum of the 23rd solar cycle~\citep{Hathaway15, Kitiashvili20}) from NASA's Advanced Composition Explorer (ACE) spacecraft, utilizing interplanetary magnetic field measurements from the Magnetometer~\citep[MAG;][]{Smith98} and solar wind velocity and proton density measurements from the Solar Wind Electron, Proton, and Alpha Monitor~\citep[SWEPAM;][]{McComas98}. The original magnetic field data at 1-second resolution~\citep{Smith22} is downsampled to 1-minute cadence, and density and velocity data at 64-second resolution~\citep{McComas22} are upsampled with linear interpolation to match the 1-minute cadence. Missing measurements are marked as ``NaN''. The resulting 12-year synchronized time series of magnetic field, velocity, and density data are used to compute the correlation functions described below. 

The present study focuses on the component-wise and trace autocorrelations of interplanetary magnetic field $B_i$, $i=\mathrm{R}, \mathrm{T}, \mathrm{N}$ (in the Radial-Tangential-Normal reference frame), which are then compared with the autocorrelation of the proton number density $n$. The two-point autocorrelation function, defined as
\begin{equation}
    R_{i}(\tau) = \langle B_i(t) B_i(t+\tau) \rangle - \langle B_i(t) \rangle \langle B_i(t+\tau)\rangle,
\label{eq:correlation}
\end{equation}
is expressed in terms of ensemble averaged quantities and $\tau$ represents the temporal lag. In practice we proceed using the {\it Blackman–Tukey} algorithm~\citep{Blackman58, Matthaeus82}, in which $\langle \cdots \rangle$ is interpreted as averaging over the entire duration of the specified data~\citep[see][for more details]{Wang24_density}. 
The temporal lag $\tau$ spans the range from zero to $T$, where $T$ is set to 0.3 times the data duration to ensure sufficient statistical weight on each correlation value.

For stationary data, Eq.~\ref{eq:correlation} is independent of the origin of the time coordinate ($t$) given sufficiently long averaging duration.
Therefore, using the property of Reynolds averages,
$\langle \langle B_i \rangle \rangle = \langle B_i\rangle$, Eq.~\ref{eq:correlation} reduces to the conventional form, $R_i(\tau) = \langle b_i(t) b_i(t+\tau) \rangle$, where $b_i(t) = B_i(t) - \langle B_i(t) \rangle$ represents the fluctuation~\citep{Germano92}. 

The component-wise magnetic field correlations are used in their unnormalized form, while the correlation trace is normalized as
\begin{equation}
    \hat{R}(\tau) = \frac{R_\mathrm{R}(\tau) + R_\mathrm{T}(\tau) + R_\mathrm{N}(\tau)}{R_\mathrm{R}(0) + R_\mathrm{T}(0) + R_\mathrm{N}(0)}.
\label{eq:correlation_normalized}
\end{equation}

We construct and analyze five variants of the decade-long magnetic field data: (1) the full dataset, (2) slow wind ($\leq \unit[400]{km/s}$ identified on a 1-minute basis from the corresponding velocity data) interleaved with gaps, (3) fast wind ($\geq \unit[500]{km/s}$) similarly identified, (4) a 4-year subset of the solar minimum, and (5) of the solar maximum. The averaging duration (including gaps) is 12 years for the first three cases and 4 years for the latter two. These data gaps are not expected to affect the correlation and spectral behaviors~\citep{Dorseth24}.

\begin{figure}
\centering
    \includegraphics[angle=0,width=0.85\columnwidth]{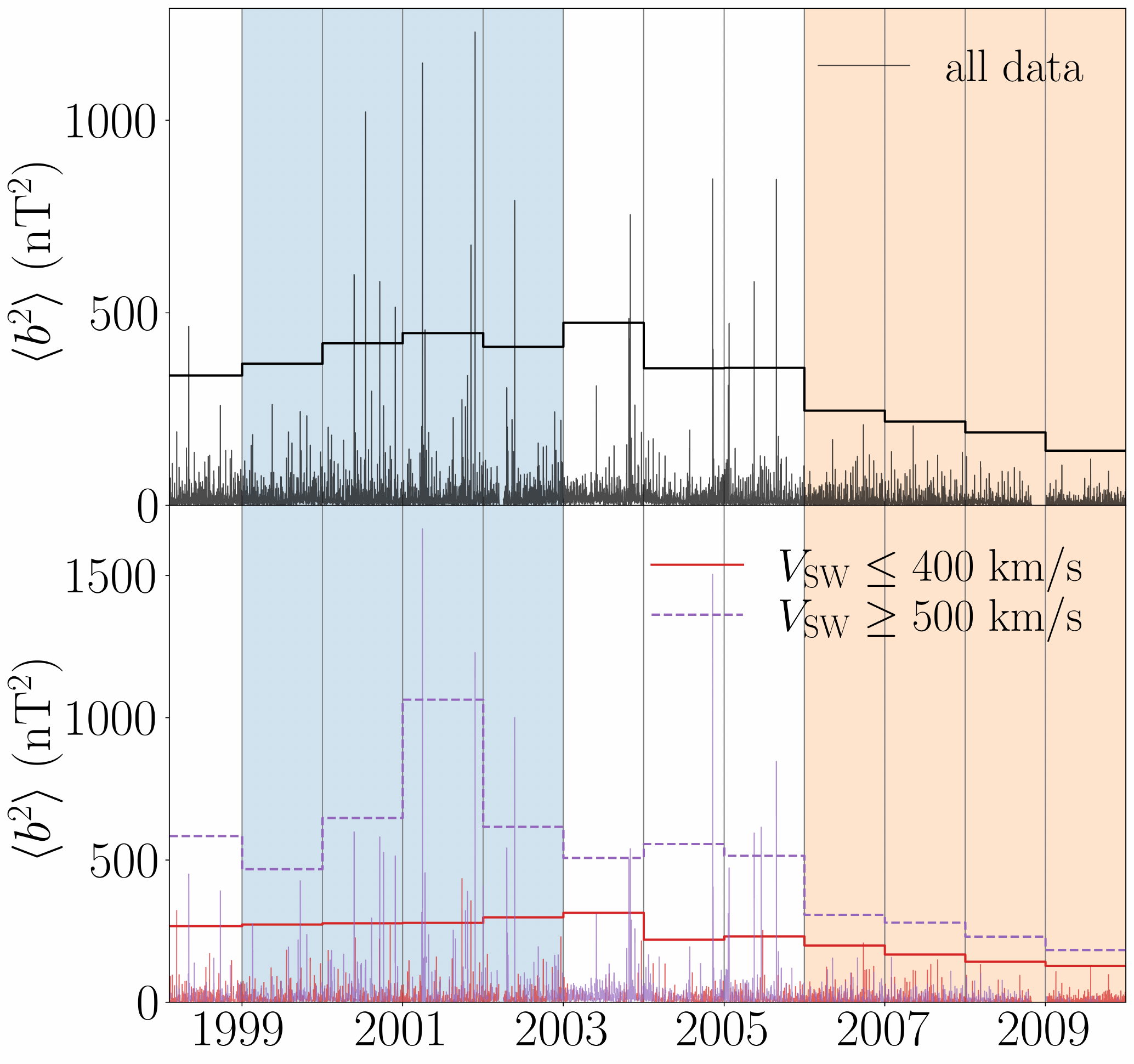}
    \caption{Magnetic field fluctuation variance $\langle b^2 \rangle = \langle b_\mathrm{R}^2 + b_\mathrm{T}^2 + b_\mathrm{N}^2 \rangle$ over 12 years (top panel), and separately for fast and slow wind (bottom panel). Piece-wise constant horizontal bars represent $\langle b^2 \rangle$ averaged over each year, then multiplied by a factor of 20 for visualization. Shaded blue and orange regions indicate the spans of solar maximum and minimum subsets, respectively.}
\label{fig:bvar}
\end{figure}

To identity the 4-year subsets of solar minimum and maximum phases, we consult the sunspot number record, the standard indicator of solar activity~\citep{Clette15}. The chosen solar maximum subset spans years 1999 to 2002, and the solar minimum subset spans 2006 to 2009. To further characterize the variability of magnetic field data in these phases, we analyze the magnetic field fluctuation variance $\langle b^2 \rangle$ over time, as shown in Fig.~\ref{fig:bvar}. Here, the component-wise magnetic field fluctuation $b_i$ is first calculated using a boxcar sliding window of 6 hours~\citep[see][for details on this method]{Matthaeus82}, approximately several times the correlation times at $\unit[1]{au}$~\citep[see, e.g., ][]{Ruiz14, Cuesta22}. The fluctuation variance is given by $\langle b^2 \rangle = \langle b_\mathrm{R}^2 \rangle + \langle b_\mathrm{T}^2 \rangle + \langle b_\mathrm{N}^2 \rangle$, where $\langle \cdots \rangle$ now denotes averaging over each consecutive 6-hour interval. As expected from heightened solar activity, the fluctuation variance is evidently higher during the solar maximum than during solar minimum. More statistics of the magnetic field fluctuation variance are shown in the SI Appendix. 

The power spectral density $S(f)$, defined as the Fourier transform of the symmetrized correlation function as
\begin{equation}
    S(f) = \int_{-\infty}^\infty R(\tau) e^{-i 2\pi f \tau} d\tau,
\end{equation}
is evaluated via the discrete Fourier transform of the bounded, discretely sampled correlation function $R(\tau)$. To prevent $R(\tau)$ from being artificially treated as periodic over the finite interval $[-T, T]$, we first apply a 10\% cosine-taper window~\citep{Matthaeus82}, in which the final 10\% of the correlation values are multiplied by the factor
\begin{equation}
    \frac{1}{2} \left( 1 + \cos{\left[ \frac{\pi}{0.1 T} (\tau - 0.9 T) \right]} \right).
\end{equation}
This is followed by padding zeros at the end of $R(\tau)$, which extends its effective duration and approximates an infinite domain~\citep{Matthaeus82}. The tapering and zero-padding techniques are found to have negligible impact on the shape of the spectral density at frequencies greater than $\unit[10^{-7}]{Hz}$, and do not affect the estimation of the integrated spectrum (explained below) above $\unit[10^{-8}]{Hz}$.

The spectral density computed in this manner shows significant fluctuations, particularly at higher frequencies, a consequence of the long data interval. To mitigate these statistical fluctuations, we apply a moving-average filter three times, where at each pass, the spectral density value is updated to $0.5$ times its original value plus $0.25$ times the values at the two neighboring frequencies~\citep[][]{Matthaeus82}. 

To examine the low-frequency spectral behavior and, in particular, to probe the extent of the $1/f$ scaling regime obscured by sharp peaks at the solar rotation frequency and its superharmonics, we integrate the power spectral density over logarithmically spaced frequency bins. These frequency bins are then systematically slid across the observed frequency range so that the mean and the standard deviation of the integrated spectrum at discrete frequencies can be computed, providing both the expected shape of the integrated spectrum and the associated uncertainty.

\section{Results}
\label{sec:results}

\subsection{Magnetic correlation}

Most past work on the magnetic autocorrelation, as reviewed in the Introduction, has reported signals at 27 days and a few harmonics in the spectrum, but only a few have analyzed the correlation functions using data records long enough to probe time lags of many months~\citep{Gosling76, Mursula96, Xie17}. Here, our emphasis is on the nature of very low frequency signals, and accordingly, our methodology employs correlation analysis with lags up to several years.
We present the analysis in several 
stages. 

\begin{figure*}
\centering
    \includegraphics[angle=0,width=0.85\columnwidth]{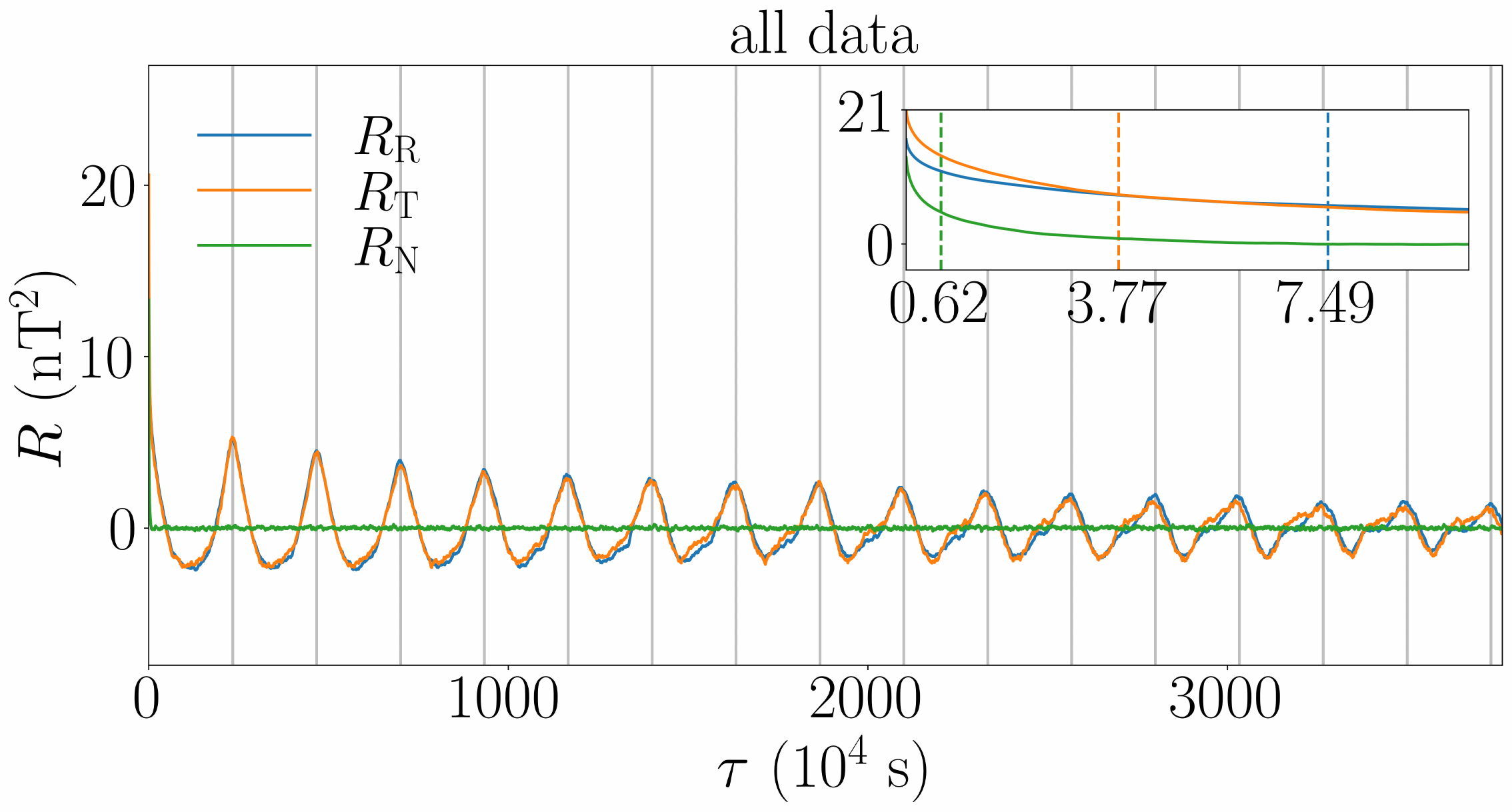}
    \includegraphics[angle=0,width=0.85\columnwidth]{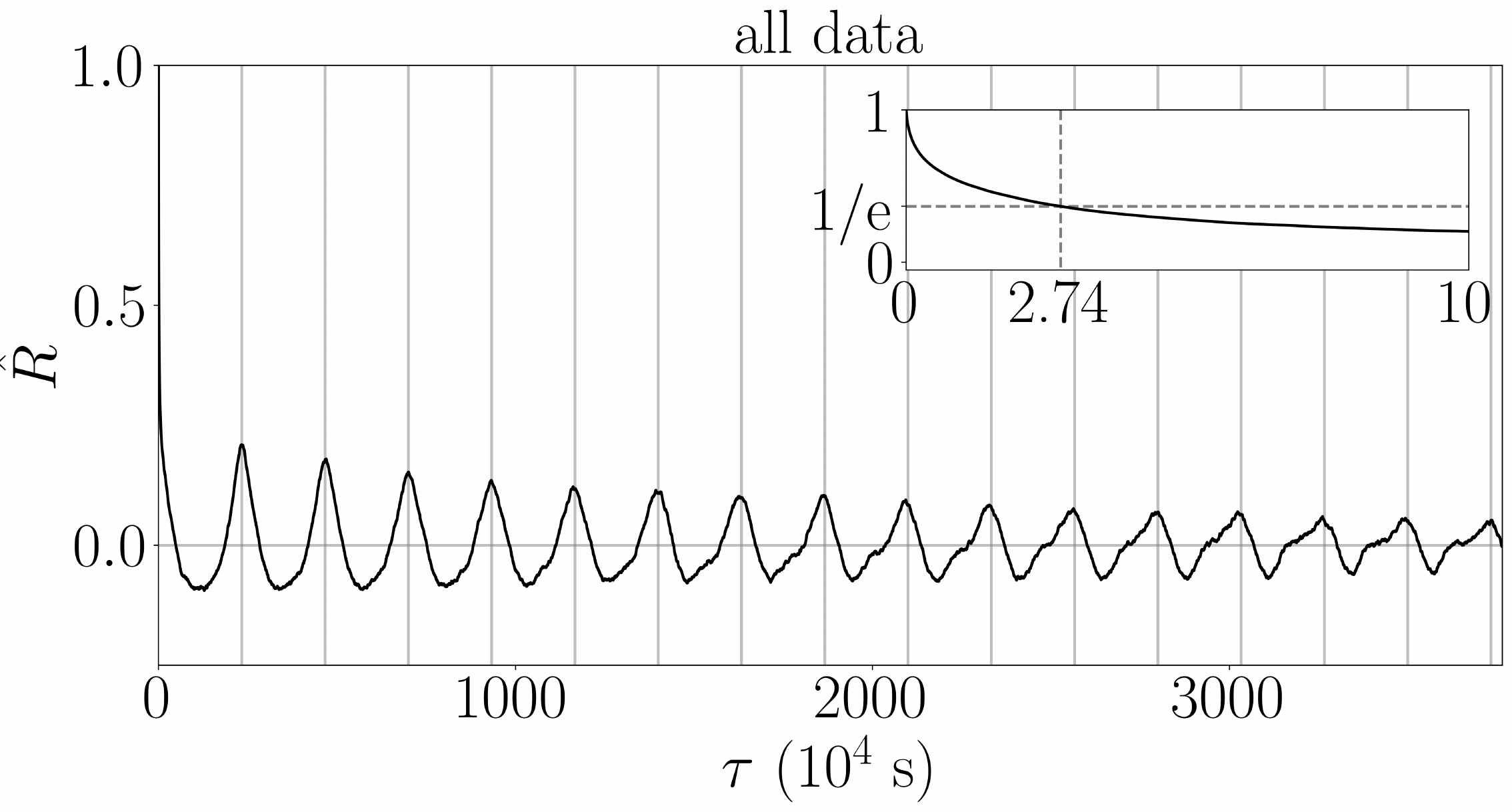}
    \caption{Magnetic field component autocorrelations (left) and normalized correlation trace (right) shown up to 1.2 year temporal lag for entire 12-year dataset. Vertical gray lines mark integer multiples of 27 days. Insets show magnified view of correlation functions up to around their correlation ($e$-folding) scales, which are indicated respectively by vertical dashed lines in matching colors. Red curve shows estimated exponential decay of local correlation maxima for each correlation trace.}
\label{fig:bcorr}
\end{figure*}

{\it Lags up to 1.2 years.}
We show the component-wise autocorrelations and the normalized correlation trace employing all 12 years of magnetic field data in Fig.~\ref{fig:bcorr}. The radial and tangential components exhibit persistent, year-long oscillations with a characteristic periodicity of approximately 27 days, corresponding to the synodic period of equatorial solar rotation. In contrast, the normal component perpendicular to the ecliptic plane shows no clear periodic behavior~\citep[consistent with][]{Matthaeus82-convergence} and decorrelates with a correlation timescale~\citep[$e$-folding time;][]{matthaeus1999correlation} of about 1.7 hours. The normalized trace, primarily influenced by the large-amplitude fluctuations in the radial and tangential components, also exhibits a 27-day periodicity, with the first several local maxima in correlation decaying approximately exponentially. We return to this point later in this section.

\begin{figure*}[t]
\centering
    \includegraphics[angle=0,width=0.85\columnwidth]{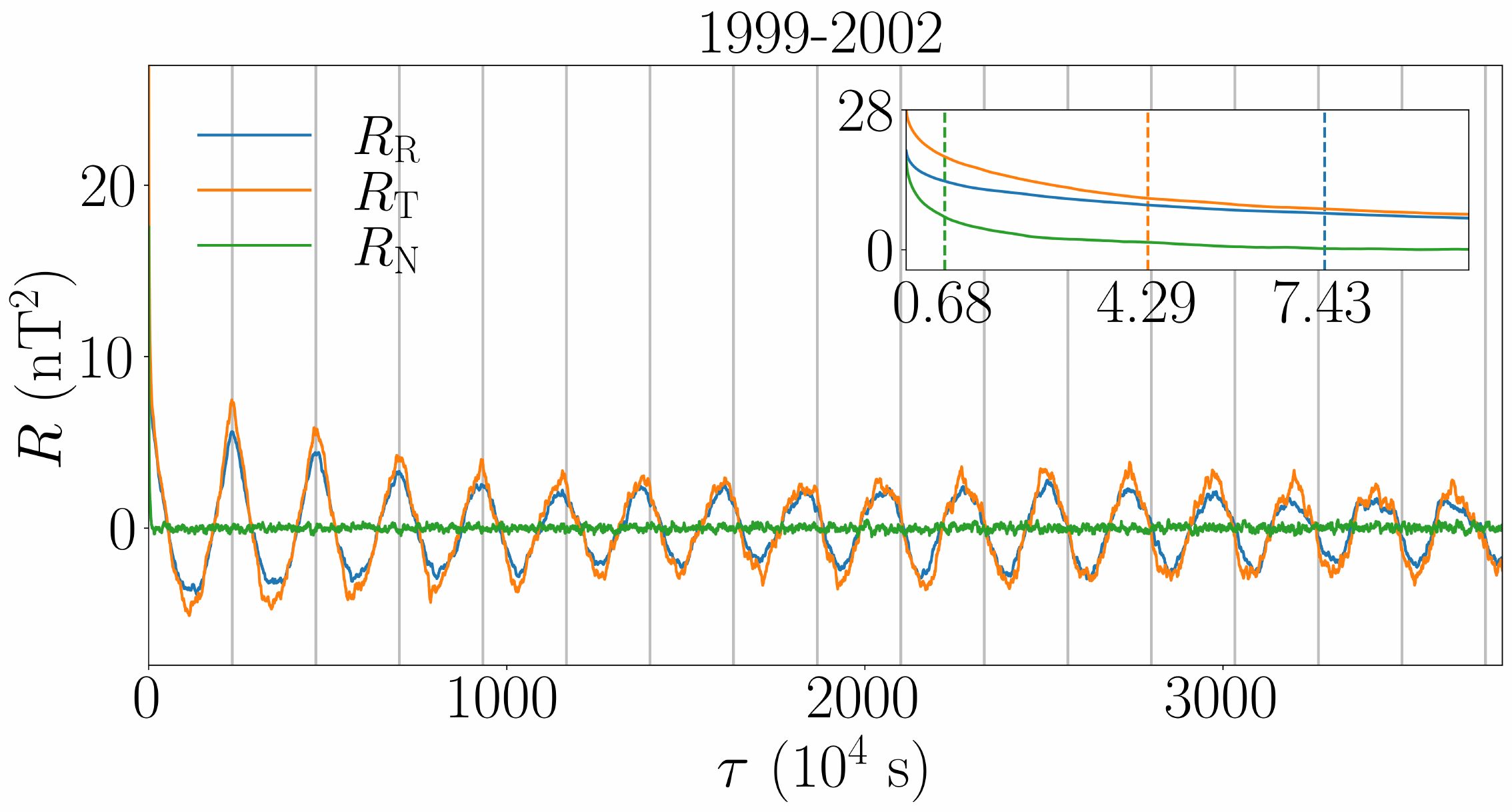}
    \includegraphics[angle=0,width=0.85\columnwidth]{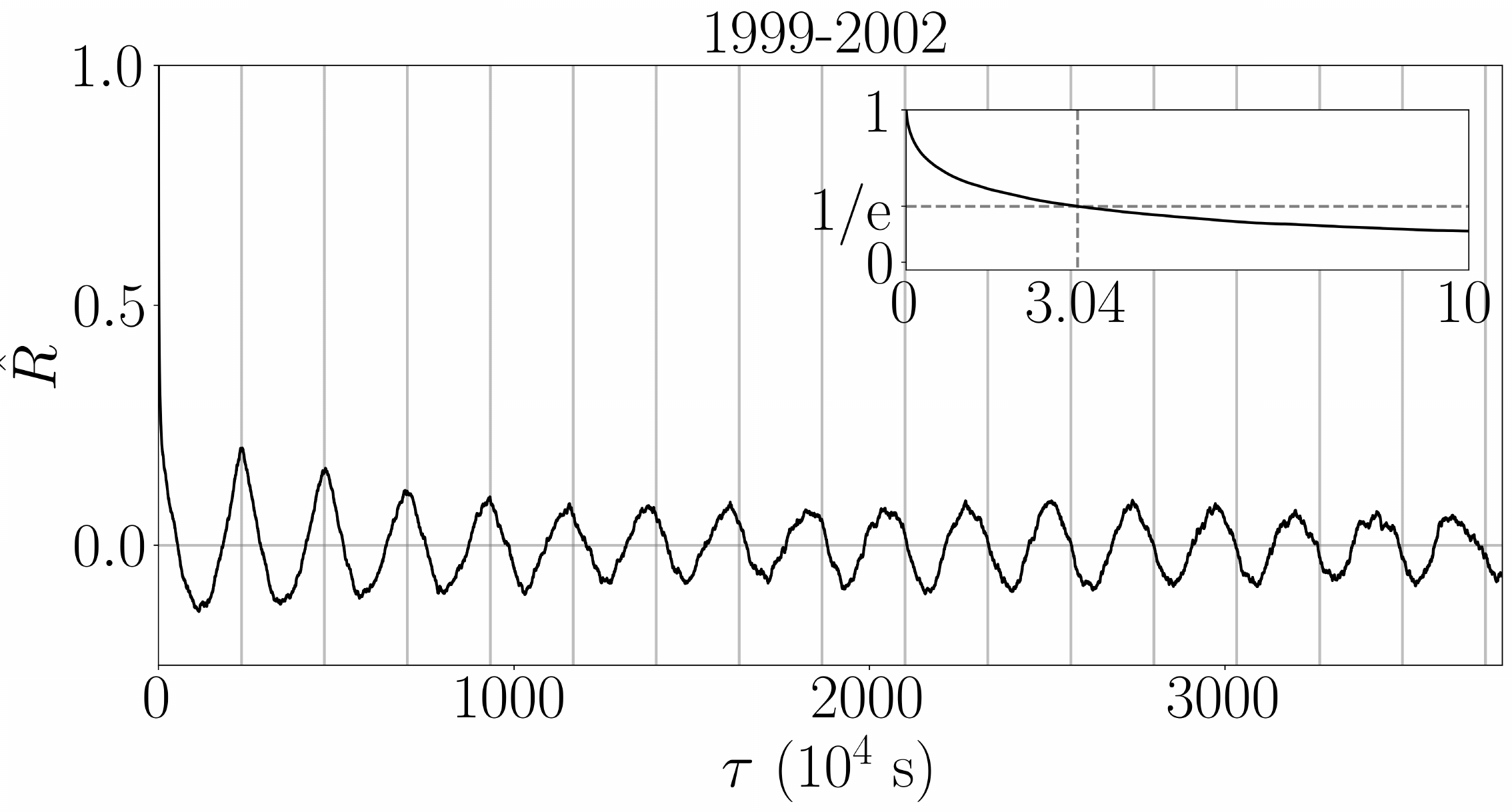}
    \includegraphics[angle=0,width=0.85\columnwidth]{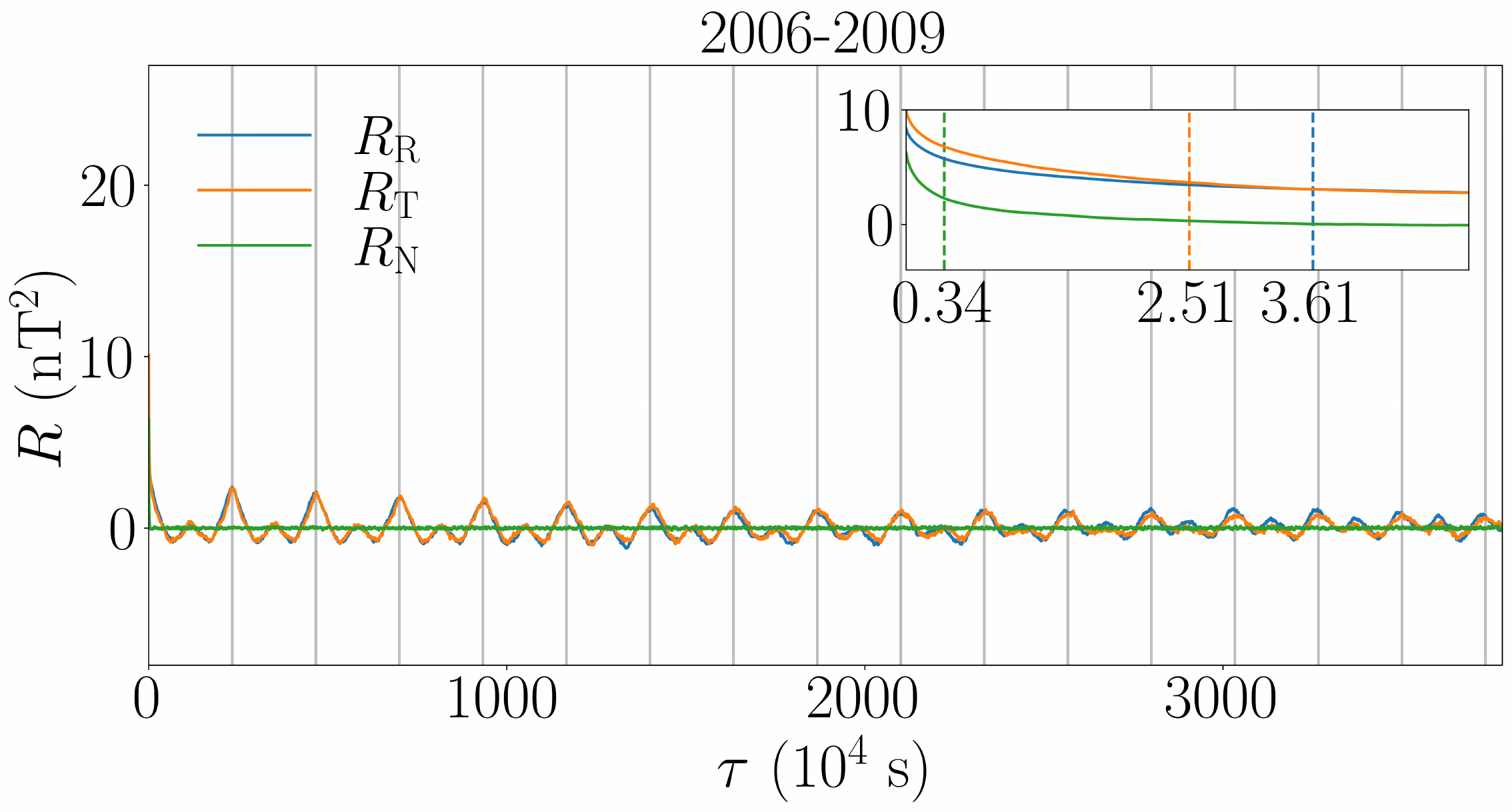}
    \includegraphics[angle=0,width=0.85\columnwidth]{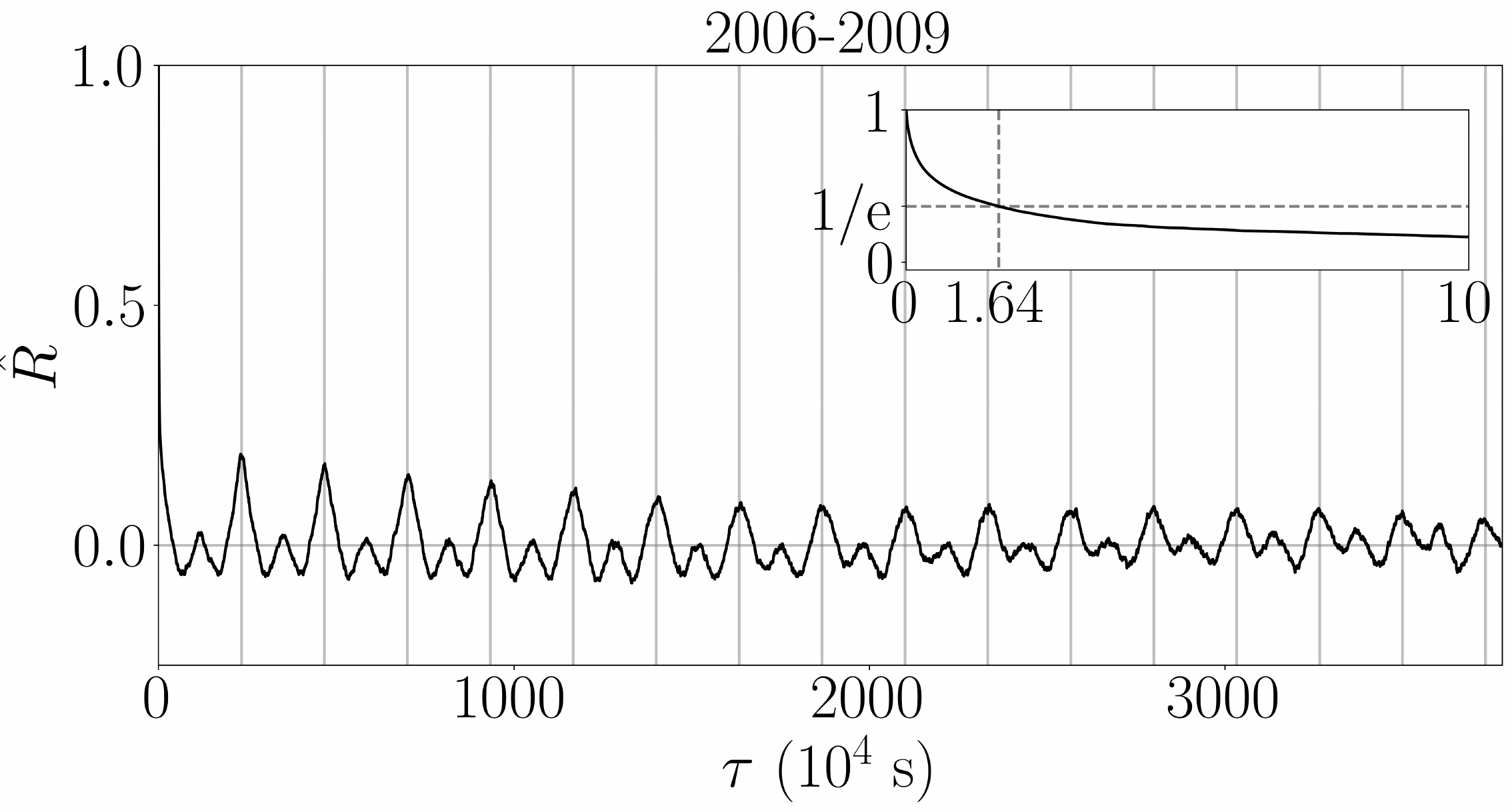}
    \caption{Same as Fig.~\ref{fig:bcorr} but for solar maximum subset from 1999 to 2002 (top), and solar minimum subset from 2006-2009 (bottom), respectively. Vertical gray lines mark integer multiples of 27 days. Insets show magnified view of correlation functions up to around their correlation ($e$-folding) scales, which are indicated respectively by vertical dashed lines in matching colors.}
\label{fig:bcorr_minmax}
\end{figure*}

The correlation functions computed separately for the solar maximum (1999-2002) and minimum (2006-2009) phases are shown in Fig.~\ref{fig:bcorr_minmax}. During solar minimum, a clear 13.5-day periodicity, corresponding to half the synodic solar rotation period, emerges alongside the primary 27-day signal. 
This feature has been reported previously as described in the Introduction. 
In contrast, the correlations during solar maximum display no distinct secondary structures. However, they begin to shift out of phase with the 27-day periodicity after approximately five solar rotations, instead exhibiting a slightly shorter period closer to around 26 days. This alludes to the possible existence of a range of periodicities, which we will examine in greater depth below.

The total energy in large-scale magnetic field fluctuations, represented by the zero-lag correlation values, is consistently greater during solar maximum and lower during solar minimum. This ordering is apparent in the inset panels of Fig.~\ref{fig:bcorr_minmax}, which emphasize behaviors at shorter time lags. Correlation times follow a similar pattern, being the longest during solar maximum and the shortest during solar minimum, with the sole exception being a slightly shorter radial correlation time during maximum compared to the full 12-year dataset. Additionally, during solar maximum, the tangential component shows stronger correlation (or anti-correlation) magnitude than the radial component, a feature less evident during solar minimum and largely averaged out over the full cycle. In both cases, the local maxima of the correlation functions do not follow a clear exponential decay trend.

\begin{figure*}
\centering
    \includegraphics[angle=0,width=0.85\columnwidth]{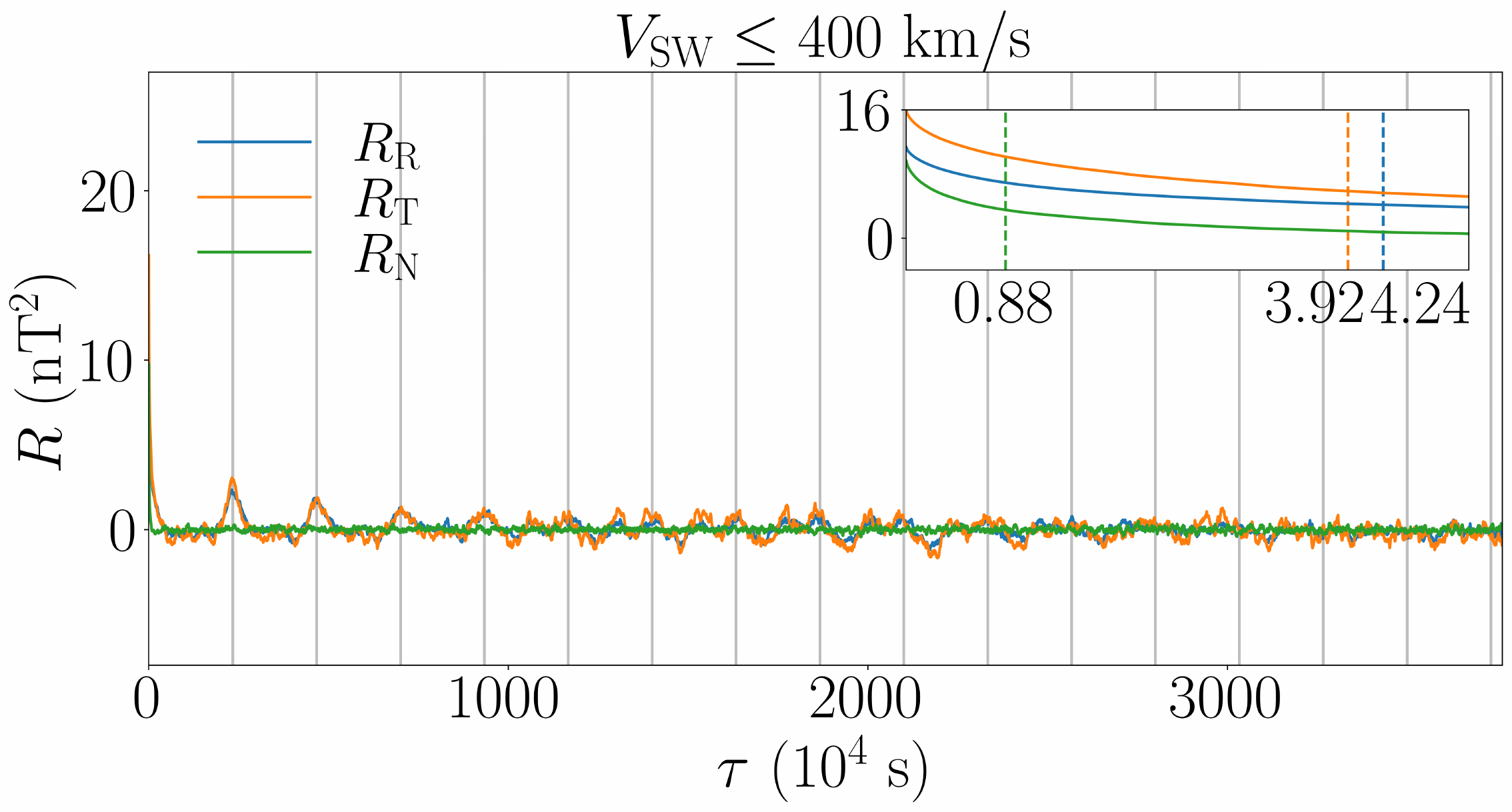}
    \includegraphics[angle=0,width=0.85\columnwidth]{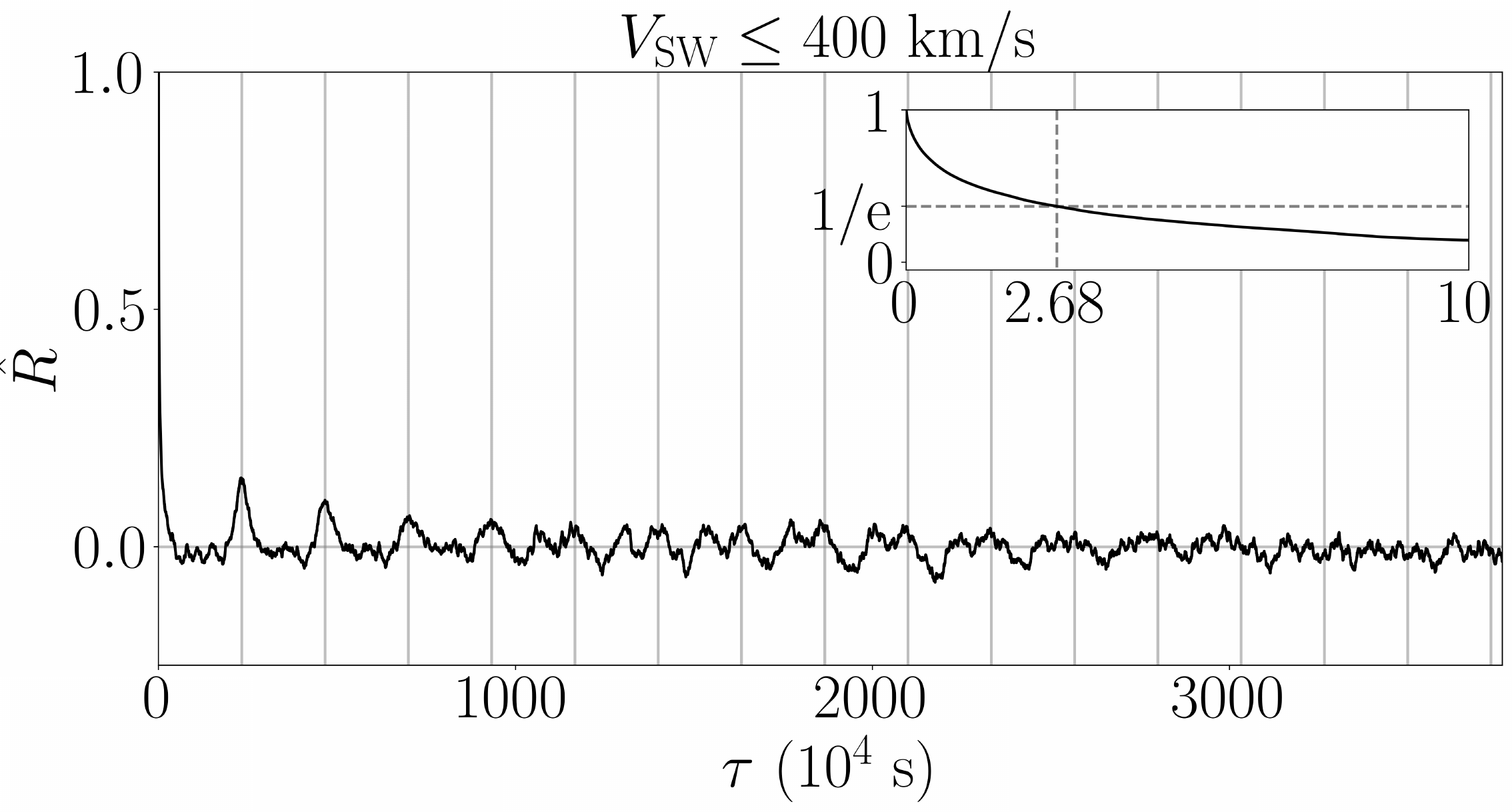}
    \includegraphics[angle=0,width=0.85\columnwidth]{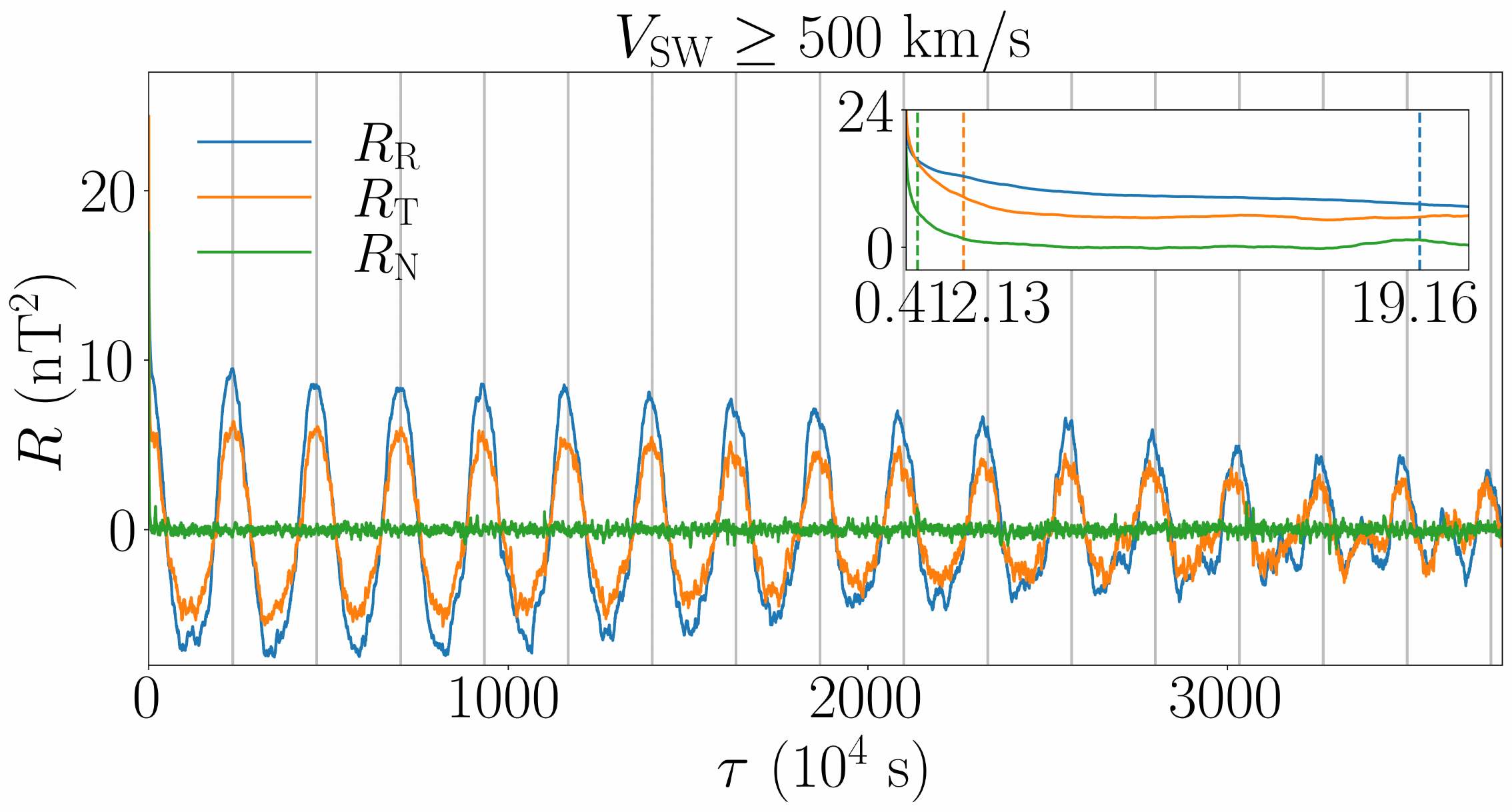}
    \includegraphics[angle=0,width=0.85\columnwidth]{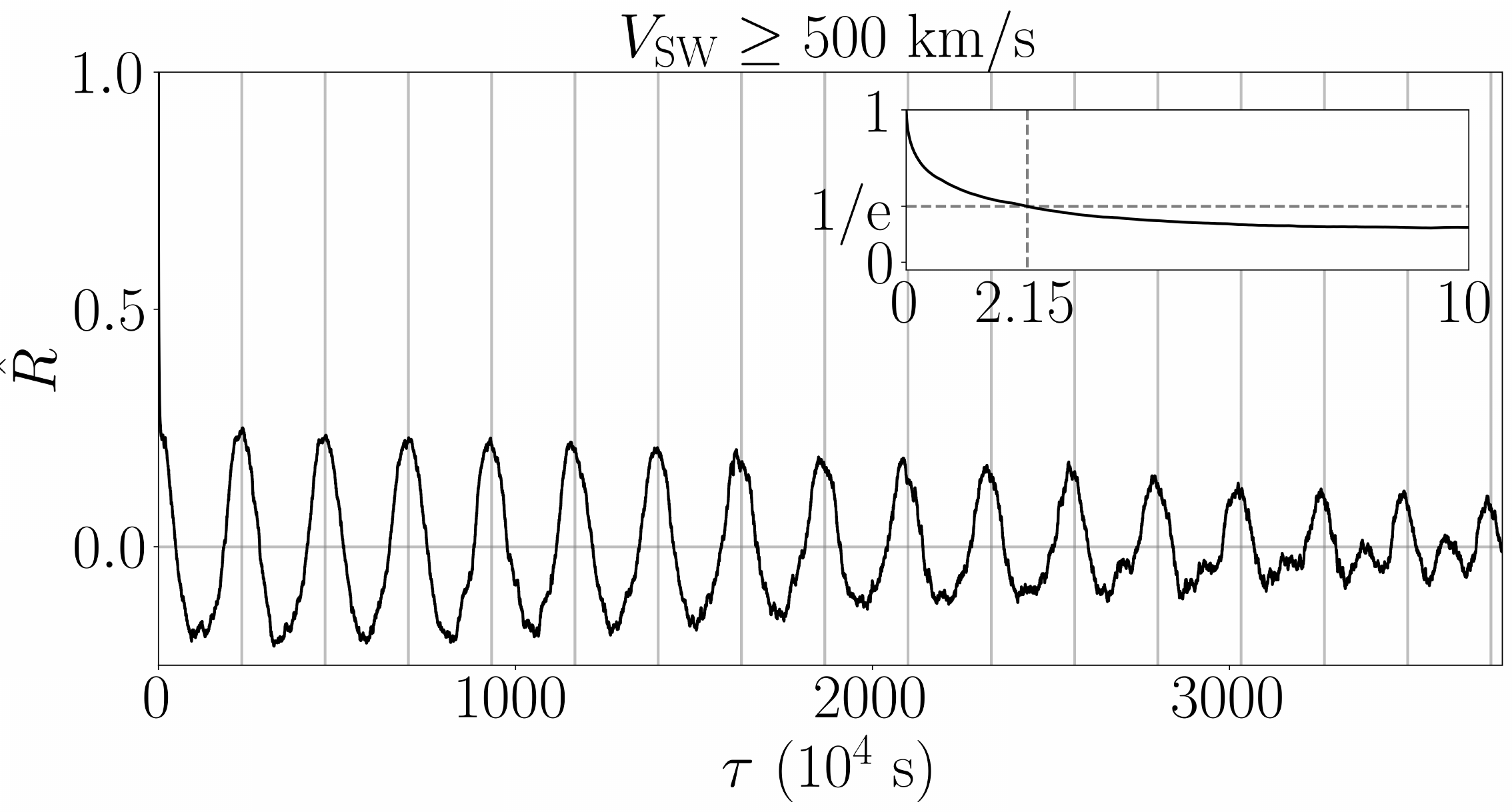}
    \caption{Same as Fig.~\ref{fig:bcorr} but for slow wind data with $V_\mathrm{SW} \leq \unit[400]{km/s}$ (top) and fast wind data with $V_\mathrm{SW} \geq \unit[500]{km/s}$ (bottom), respectively. Vertical gray lines mark integer multiples of 27 days. Insets show magnified view of correlation functions up to around their correlation ($e$-folding) scales, which are indicated respectively by vertical dashed lines in matching colors.}
\label{fig:bcorr_slowfast}
\end{figure*}

The correlation functions computed separately for the fast solar wind and the slow solar wind are shown in Fig.~\ref{fig:bcorr_slowfast}. In slow wind, the recurrence of correlation at 27 days disappears after around four solar rotations, and is overpowered in magnitude by some higher-frequency structures. At the same time, oscillations at finer scales in lag become discernible. Structures in fast wind, however, predominantly display a 27-day periodicity with 13.5-day secondary structures emerging after around 12 solar rotations. 

In comparing analyses among the slow wind, fast wind, and the full dataset, the large-scale fluctuation variance in all magnetic field components is consistently the lowest for the slow wind. Notably, slow wind has the longest normal and tangential correlation times in comparison with fast wind and the full dataset; conversely, the slow wind radial correlation time is the lowest. The fast wind correlation times exhibit the opposite trend, being the highest in the radial and the lowest in the normal and tangential directions.

{\it Lags up to 3.6 years.} 
An interesting change exists in 
the qualitative behavior of 
correlations at the shorter lags 
described above and behavior of the same functions when computed at significantly longer lags. 
When examining the correlation functions over temporal lags smaller than 1.2 years, we notice that the long-lag behavior does not follow a consistent trend. In particular, while the 12-year dataset shows an approximately exponential decay in the local correlation maxima, this trend is absent in several subsets, such as the slow wind and the 1999-2002 solar maximum. Therefore, we now 
extend the correlation computation to 3.6 years, corresponding to 30\% of the total data duration.

\begin{figure*}
\centering
    \includegraphics[angle=0,width=0.85\columnwidth]{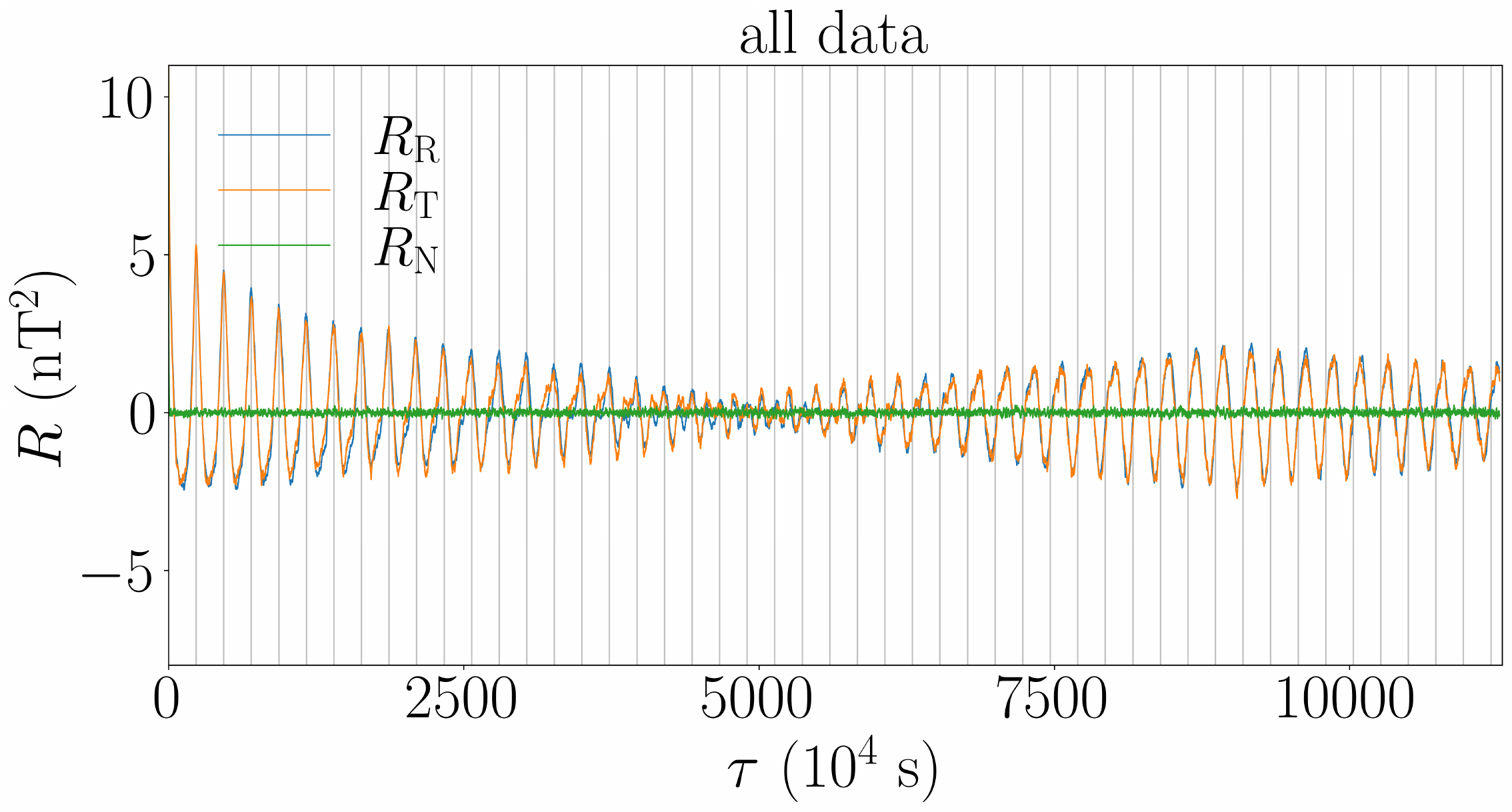}
    \includegraphics[angle=0,width=0.85\columnwidth]{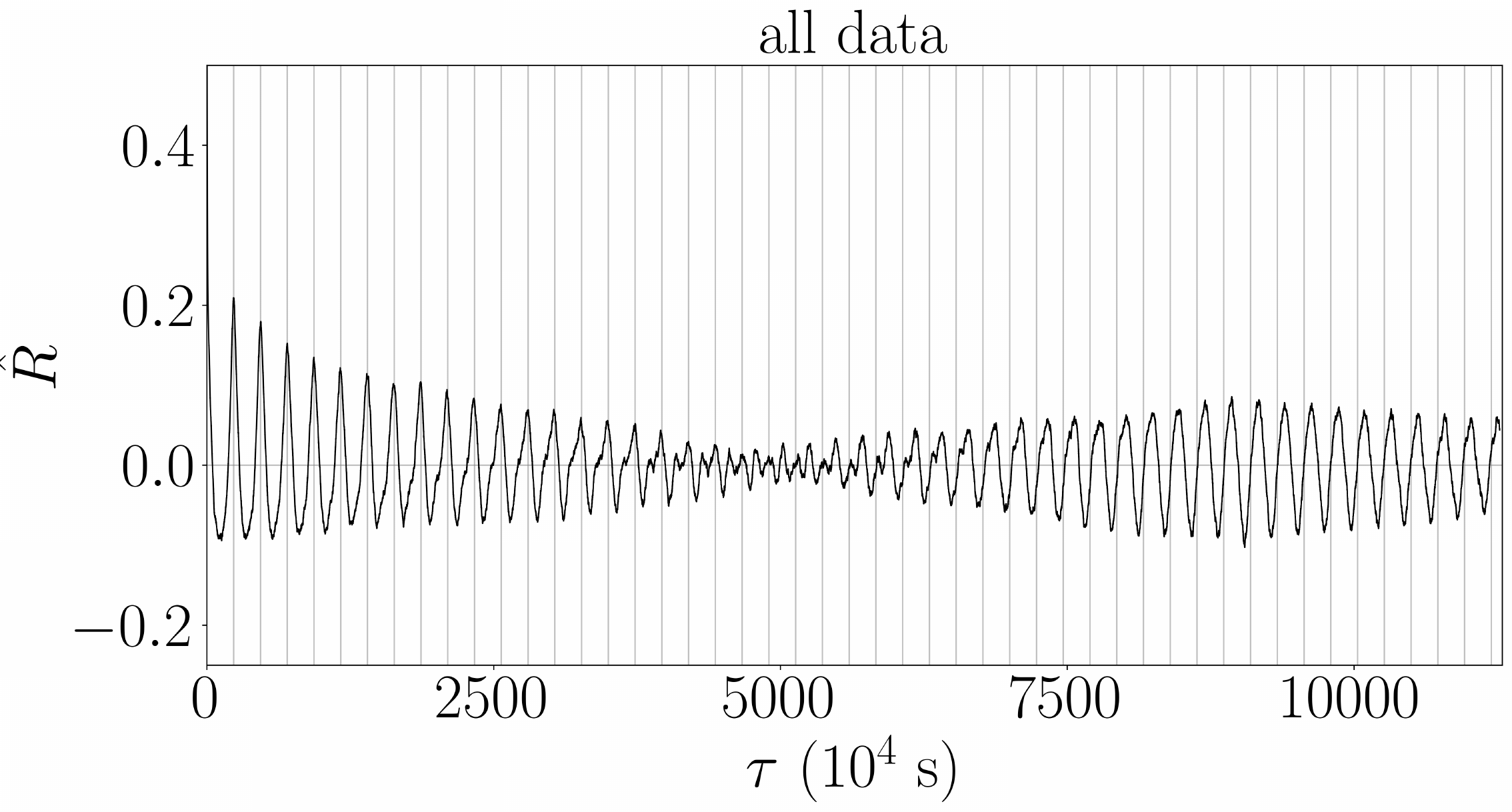}
    \includegraphics[angle=0,width=0.85\columnwidth]{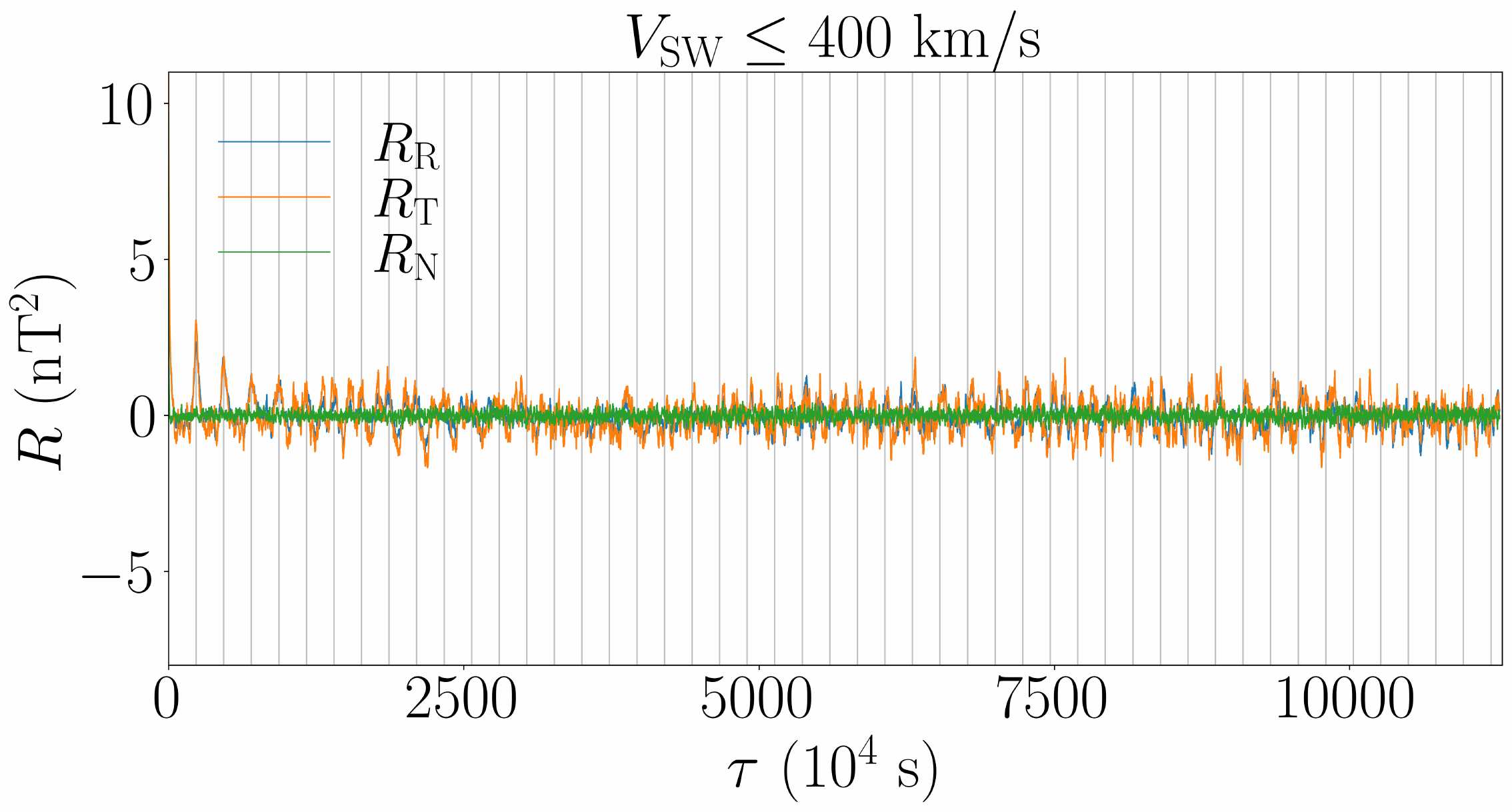}
    \includegraphics[angle=0,width=0.85\columnwidth]{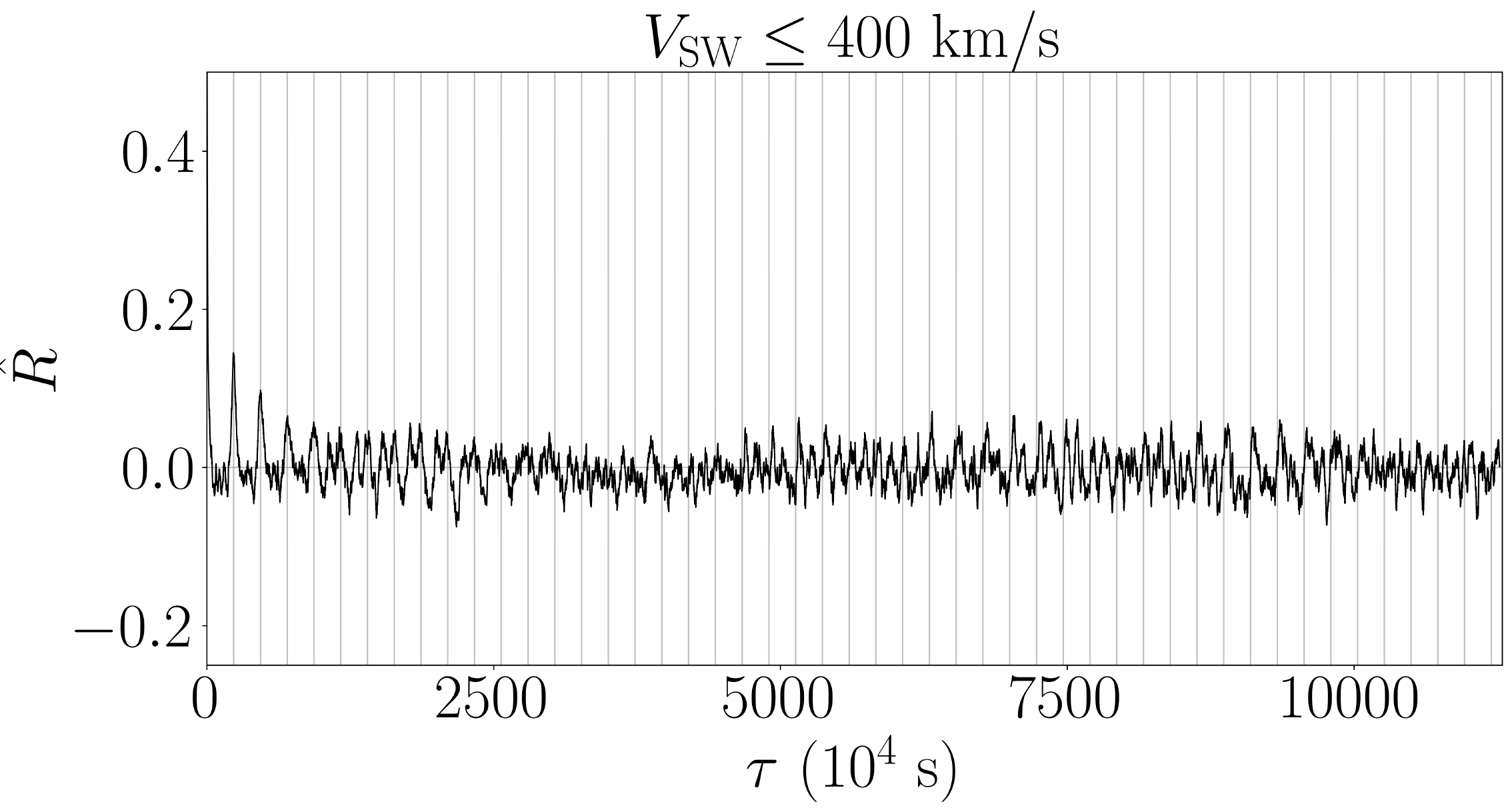}
    \includegraphics[angle=0,width=0.85\columnwidth]{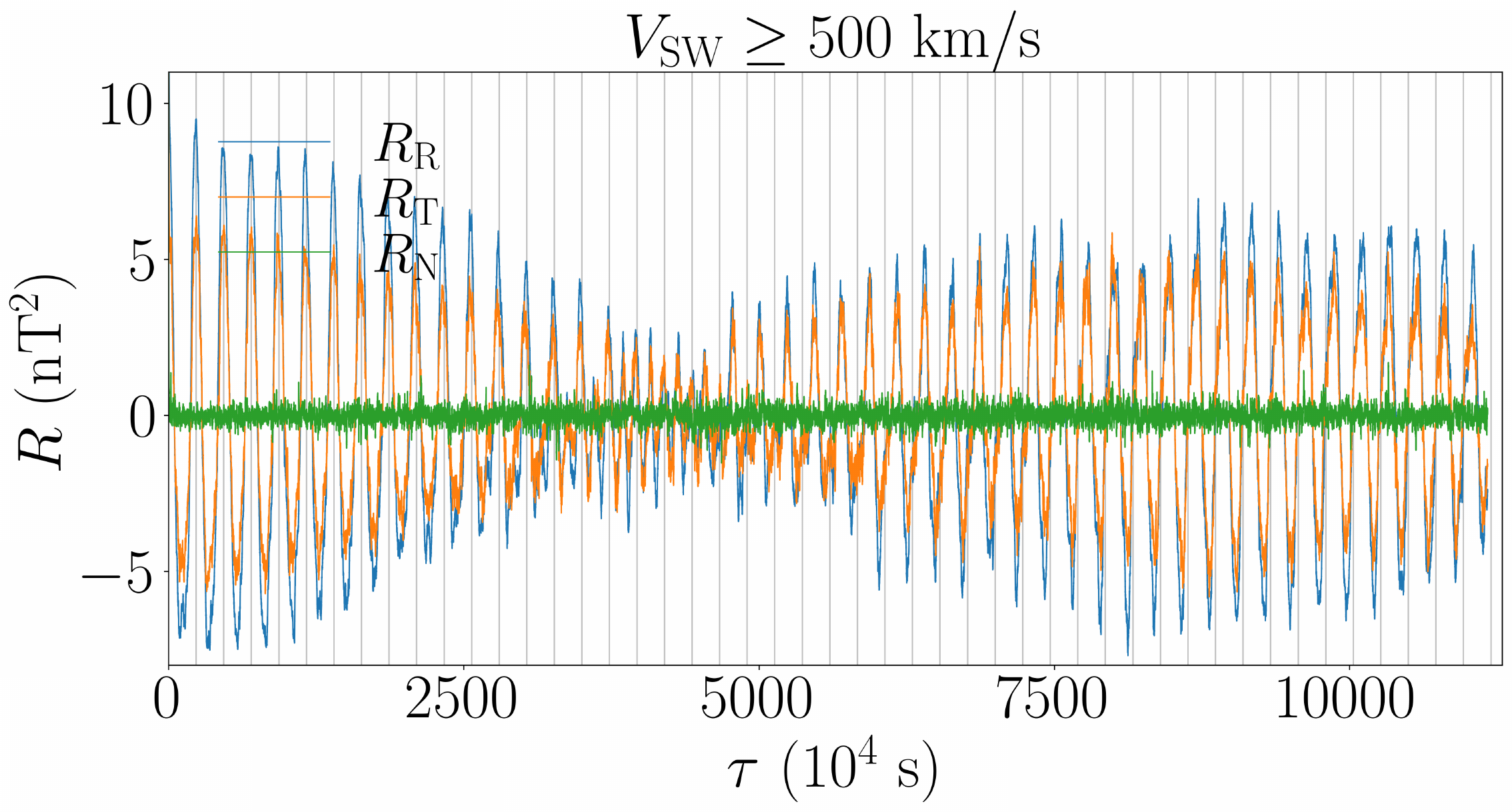}
    \includegraphics[angle=0,width=0.85\columnwidth]{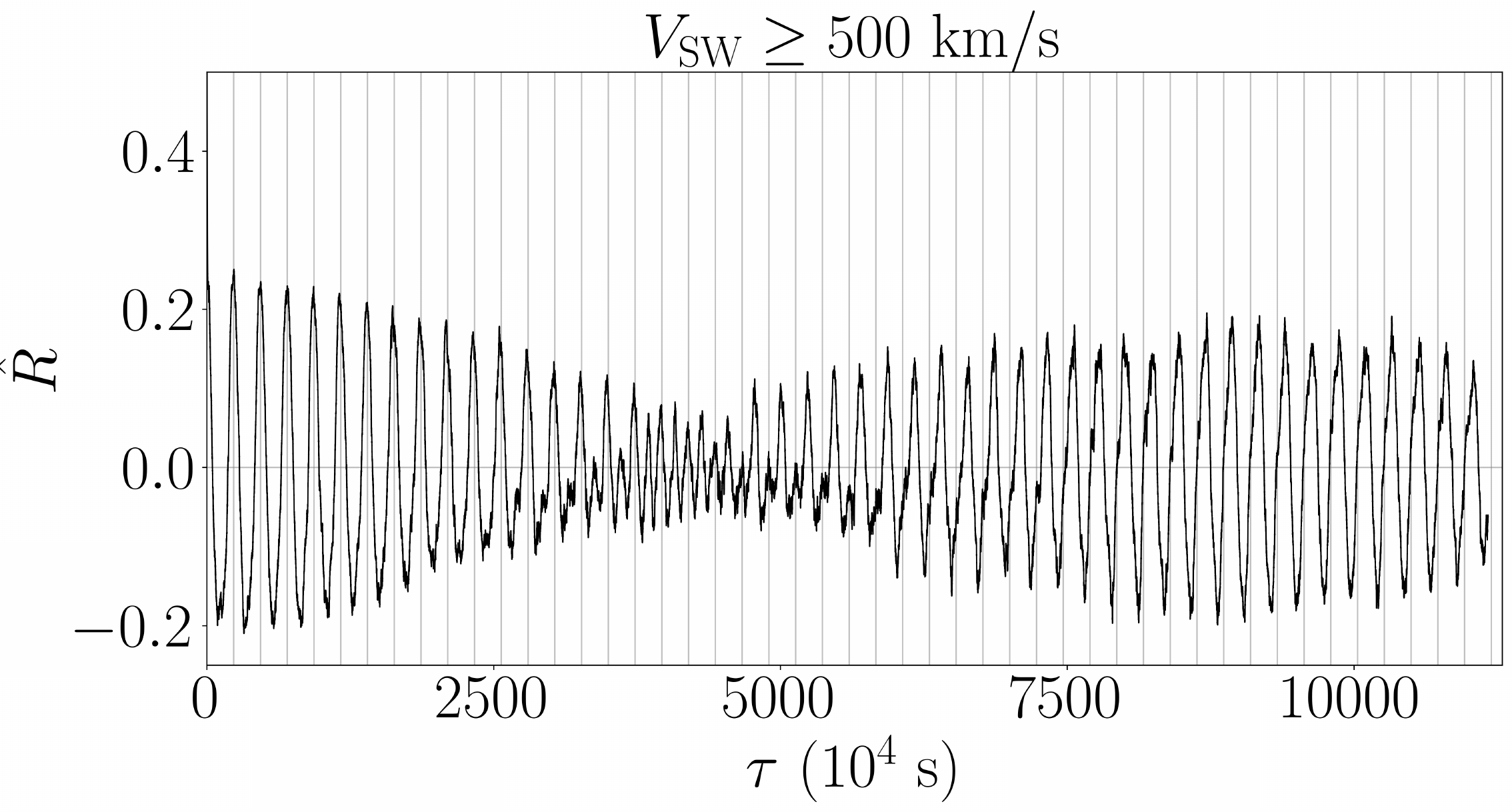}
    \caption{Magnetic field component correlations (left) and normalized correlation trace (right), same as those in Figs.~\ref{fig:bcorr} and~\ref{fig:bcorr_slowfast} but shown up to 3.6 years temporal lag, for entire 12-year data (top) and slow (middle) and fast wind (bottom). Vertical gray lines mark integer multiples of 27 days. Vertical scale is adjusted for clarity 
    and zero-lag correlation maxima are not shown.}
\label{fig:bcorr_long}
\end{figure*}

We show in Fig.~\ref{fig:bcorr_long} the extended component-wise correlations and the normalized correlation trace for the full 12-year dataset, as well as for the slow and fast wind intervals.
One prominent feature is the presence of year-long oscillation envelope superimposed on the dominant $\sim27$-day fluctuations. The configuration resembles the ``beat'' phenomenon in acoustics, where the interference of waves with closely separated frequencies produces a low-frequency modulation in the wave amplitude. In the context of the solar wind, differential solar rotation or structures in the interplanetary wind may introduce a range of periodicities near 27 days, resulting in a similar beating effect. This mechanism could account for the gradual shift in periodicity after several solar rotations as observed during solar maximum in Fig.~\ref{fig:bcorr_minmax}, as well as the erratic evolution of local correlation maxima over lag.

\subsection{Magnetic spectrum}

\begin{figure*}
\centering
    \includegraphics[angle=0,width=0.6\columnwidth]{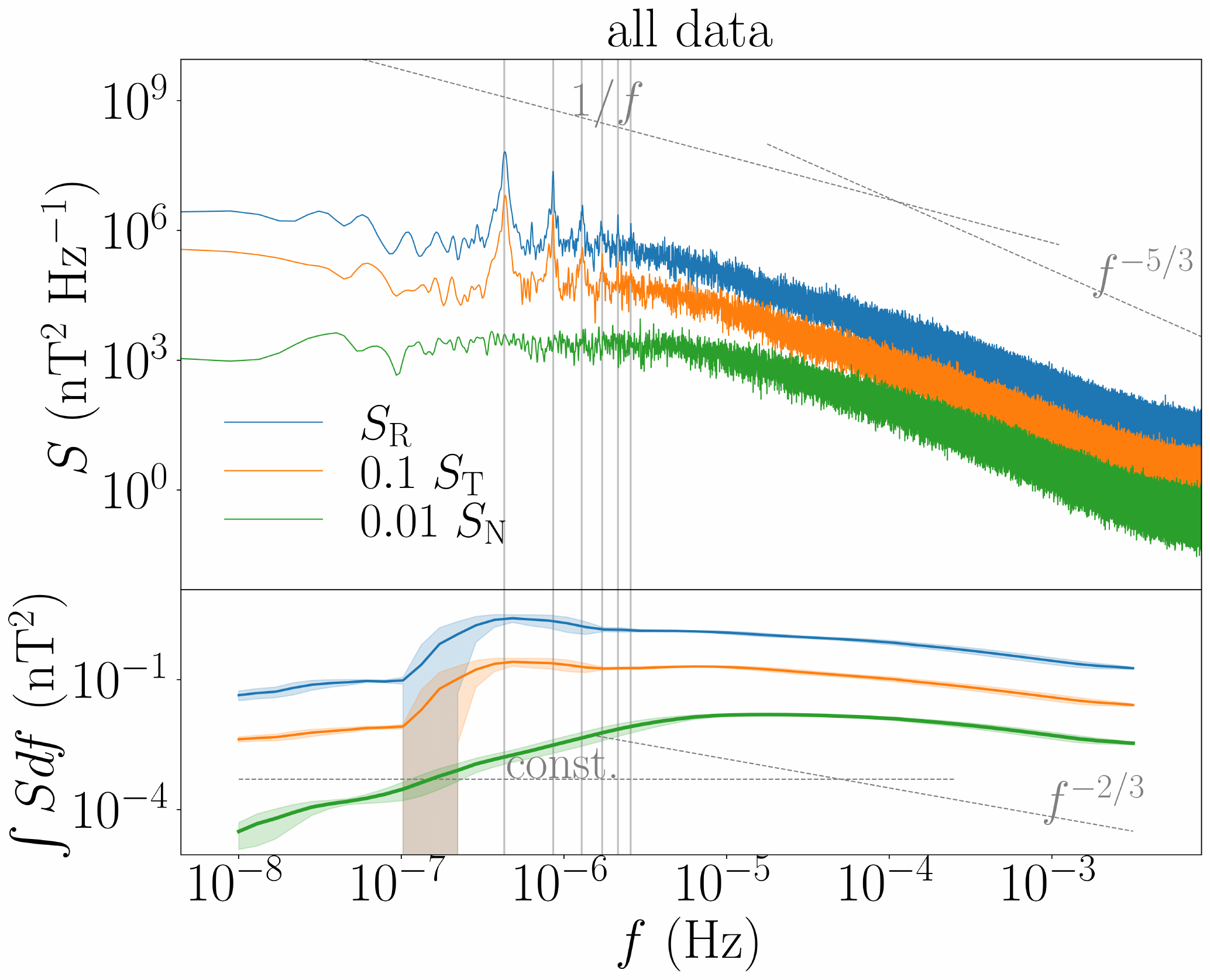}
    \includegraphics[angle=0,width=0.6\columnwidth]{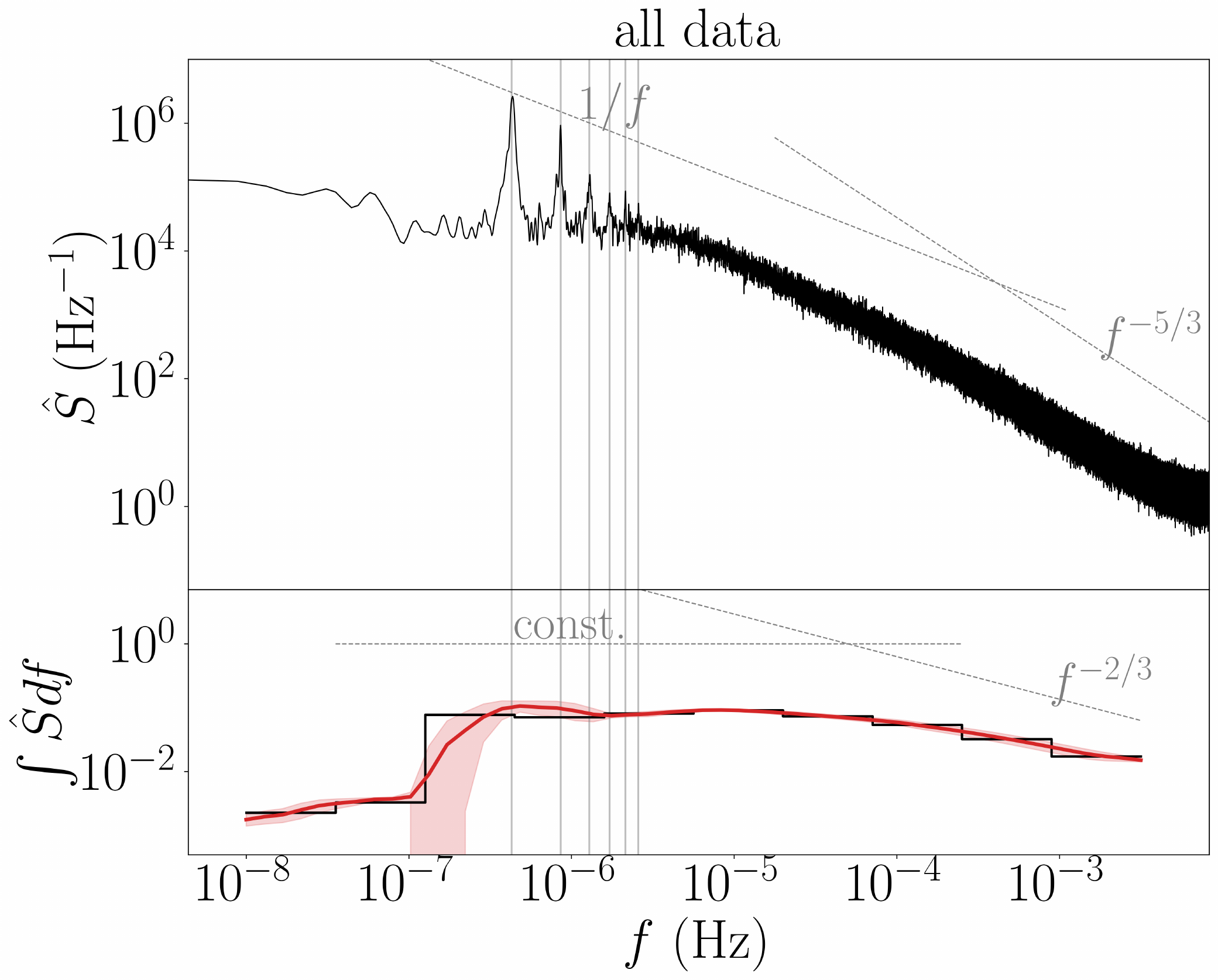}
    
    \includegraphics[angle=0,width=0.6\columnwidth]{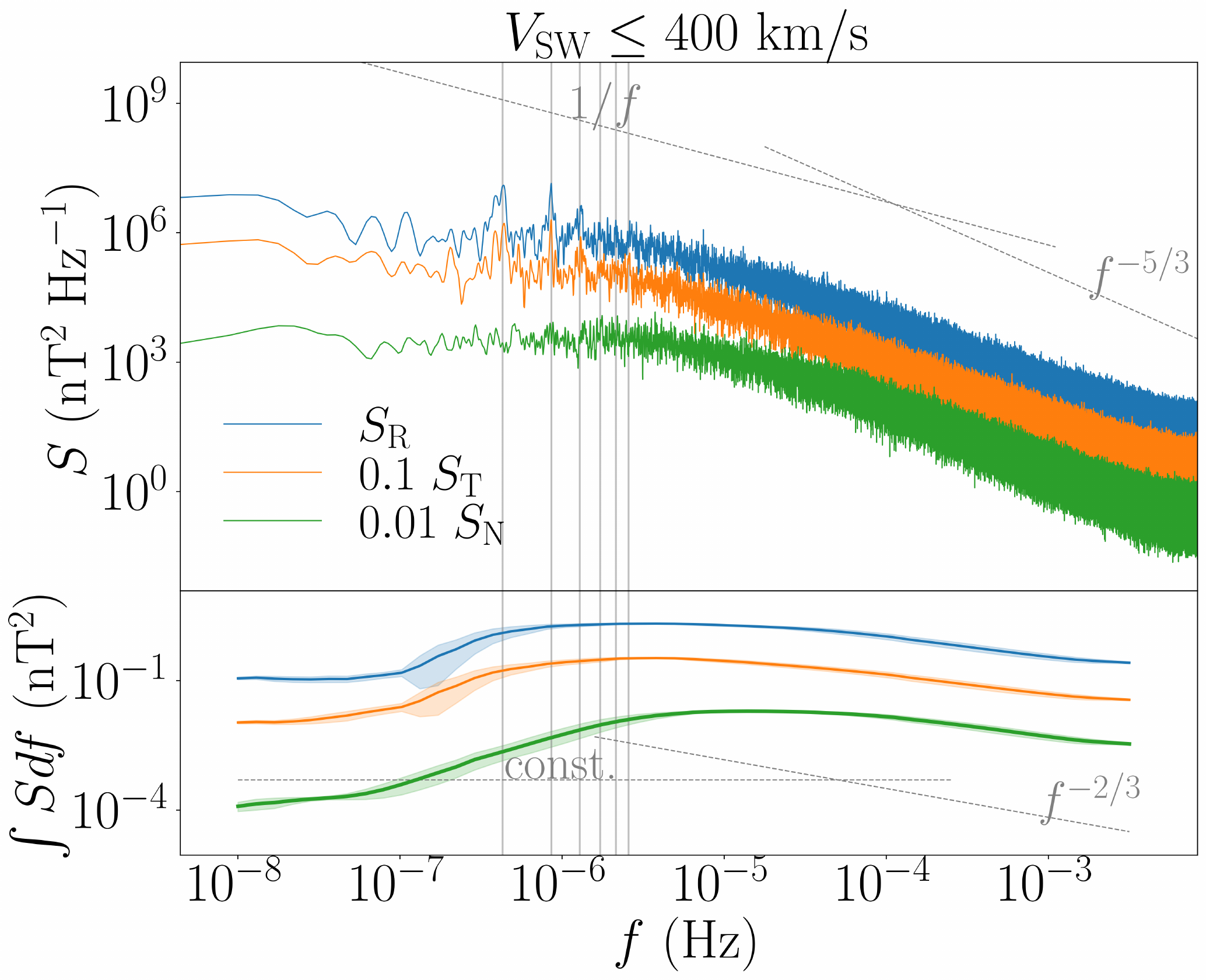}
    \includegraphics[angle=0,width=0.6\columnwidth]{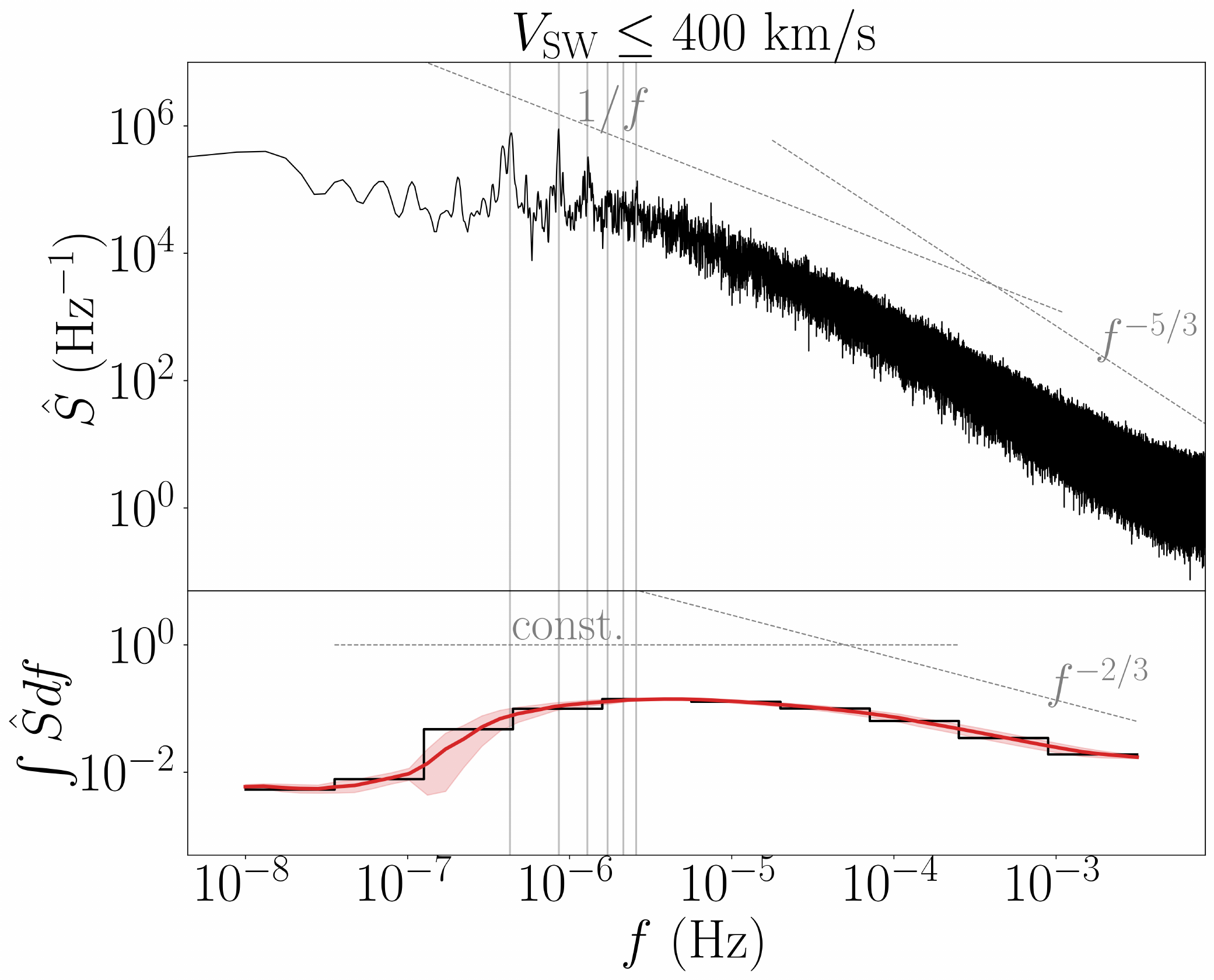}
    
    \includegraphics[angle=0,width=0.6\columnwidth]{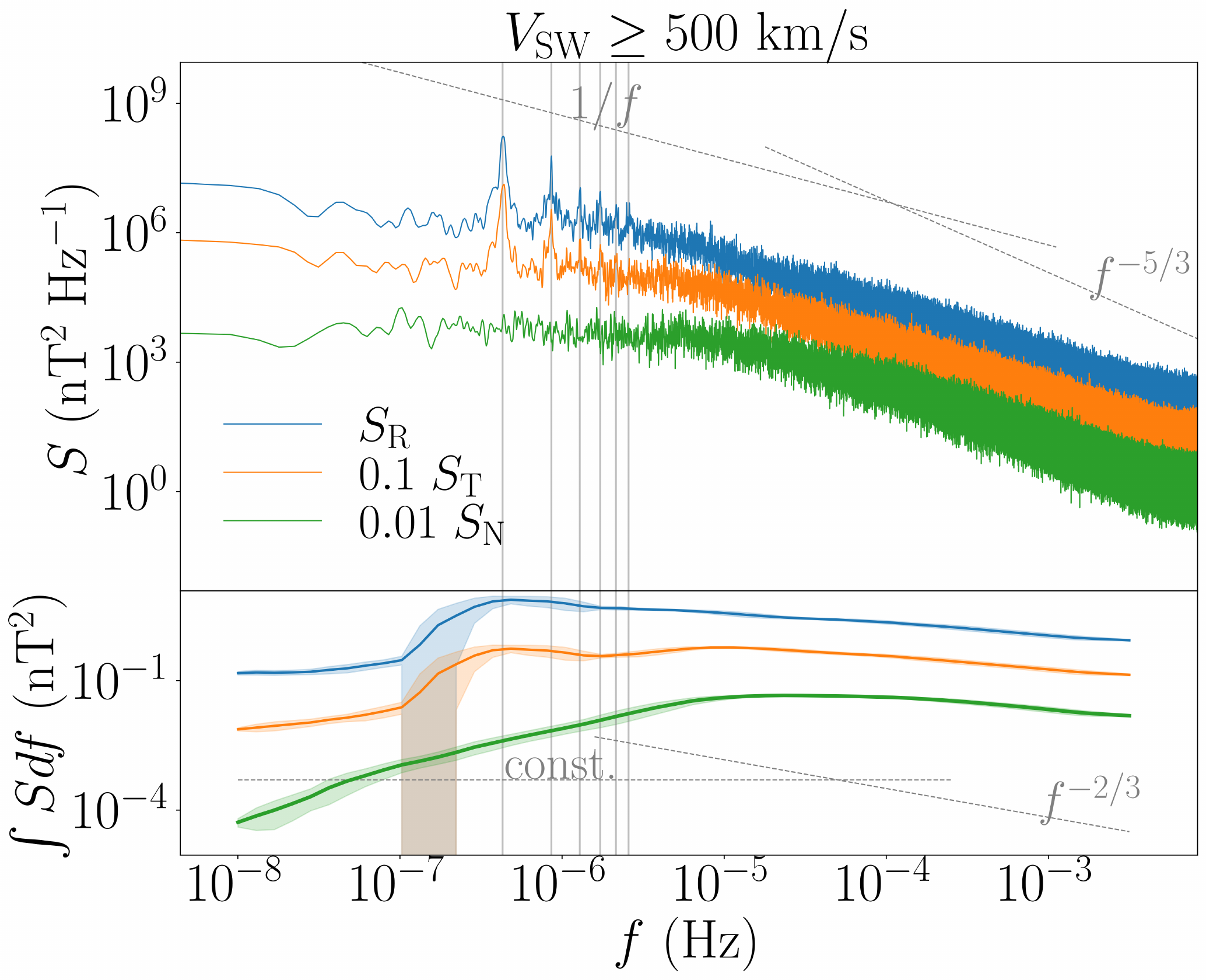}
    \includegraphics[angle=0,width=0.6\columnwidth]{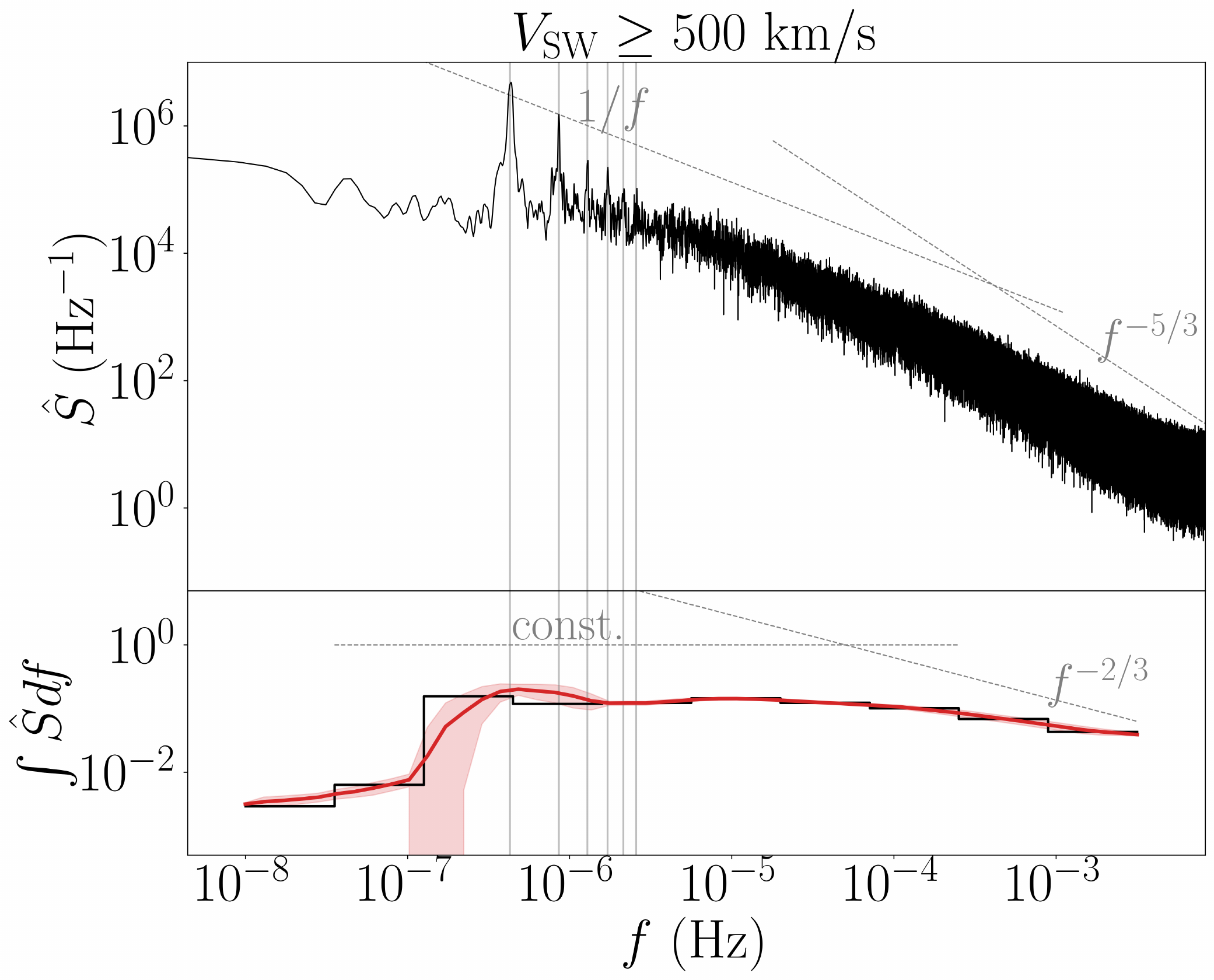}
    
    \includegraphics[angle=0,width=0.6\columnwidth]{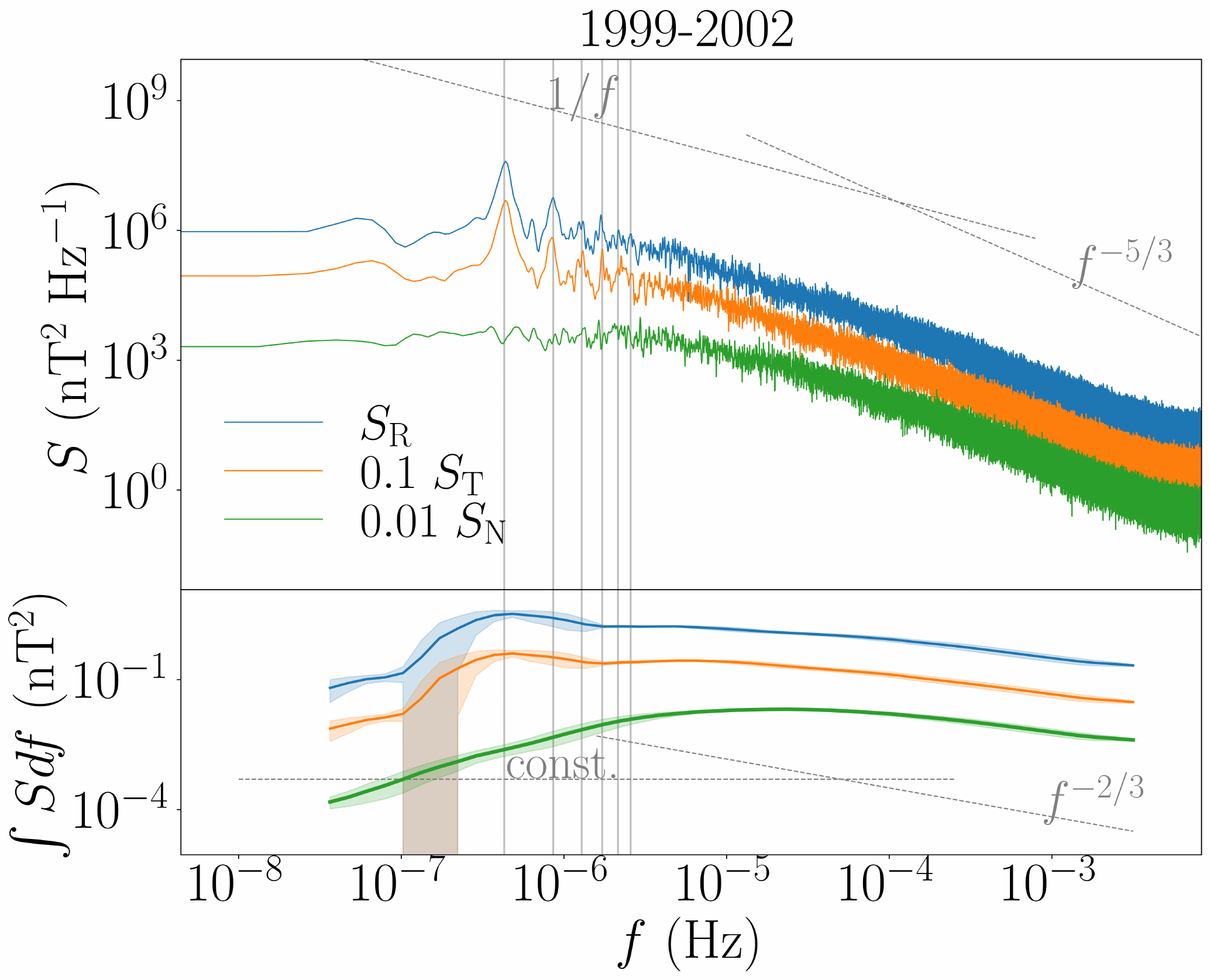}
    \includegraphics[angle=0,width=0.6\columnwidth]{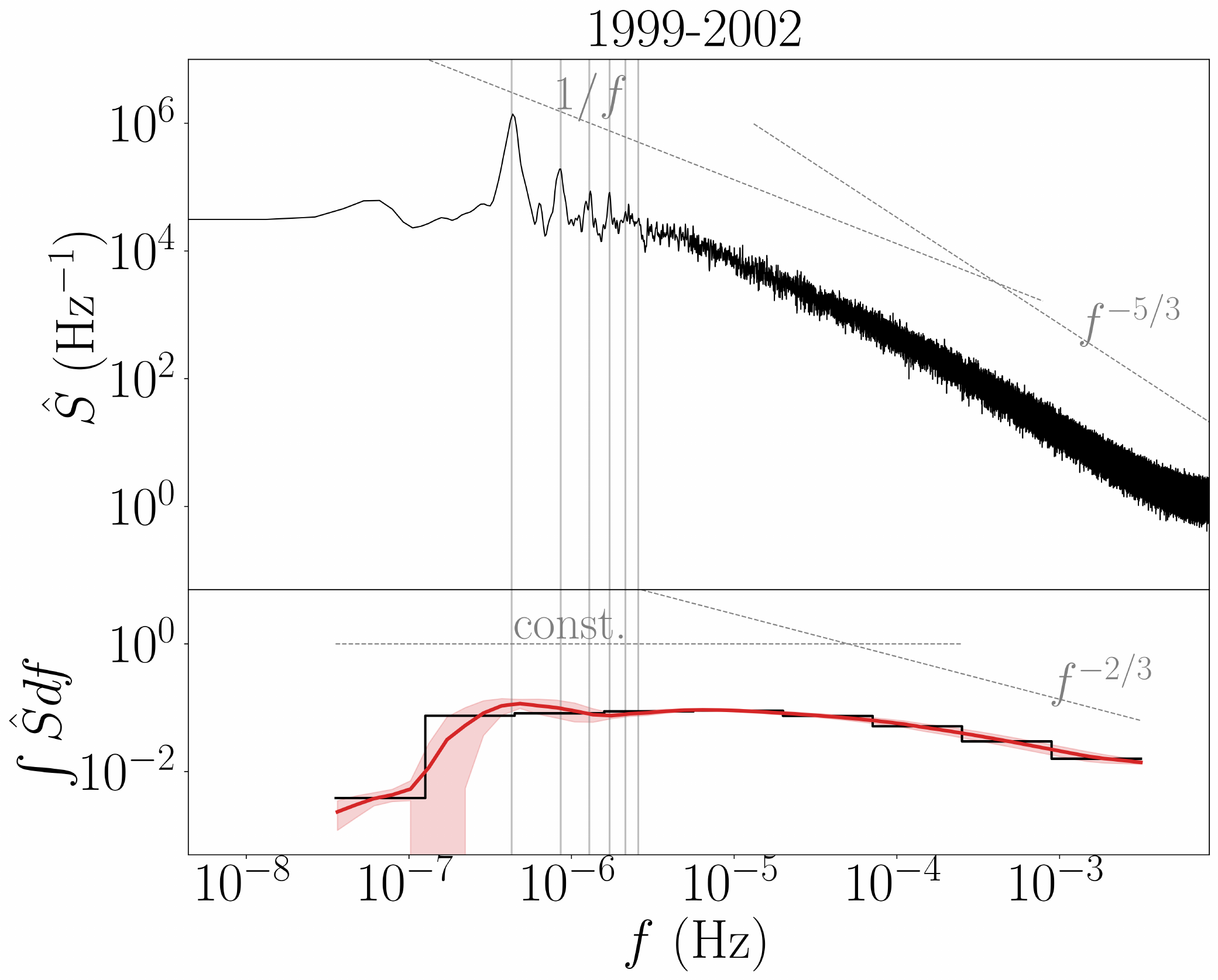}
    
    \includegraphics[angle=0,width=0.6\columnwidth]{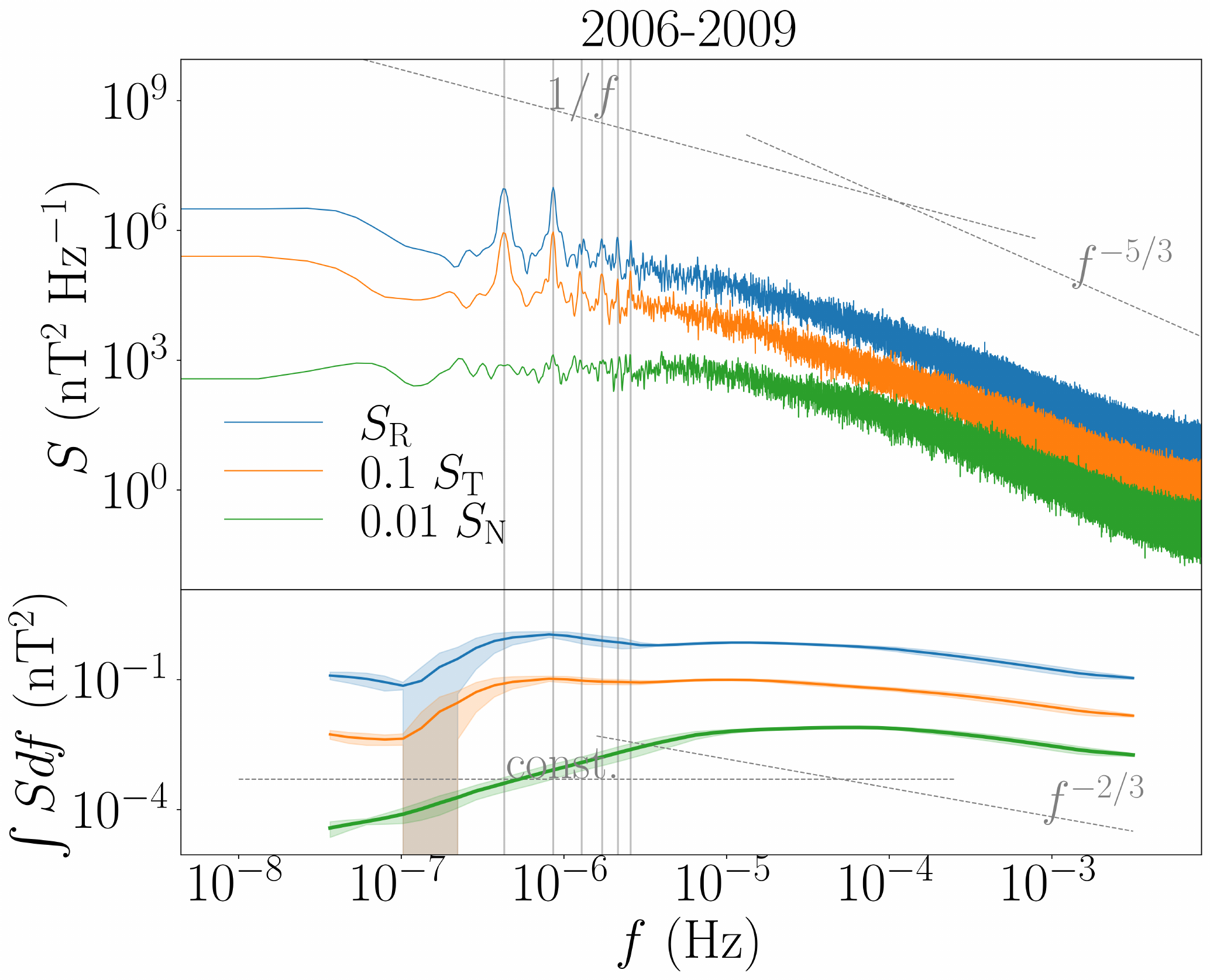}
    \includegraphics[angle=0,width=0.6\columnwidth]{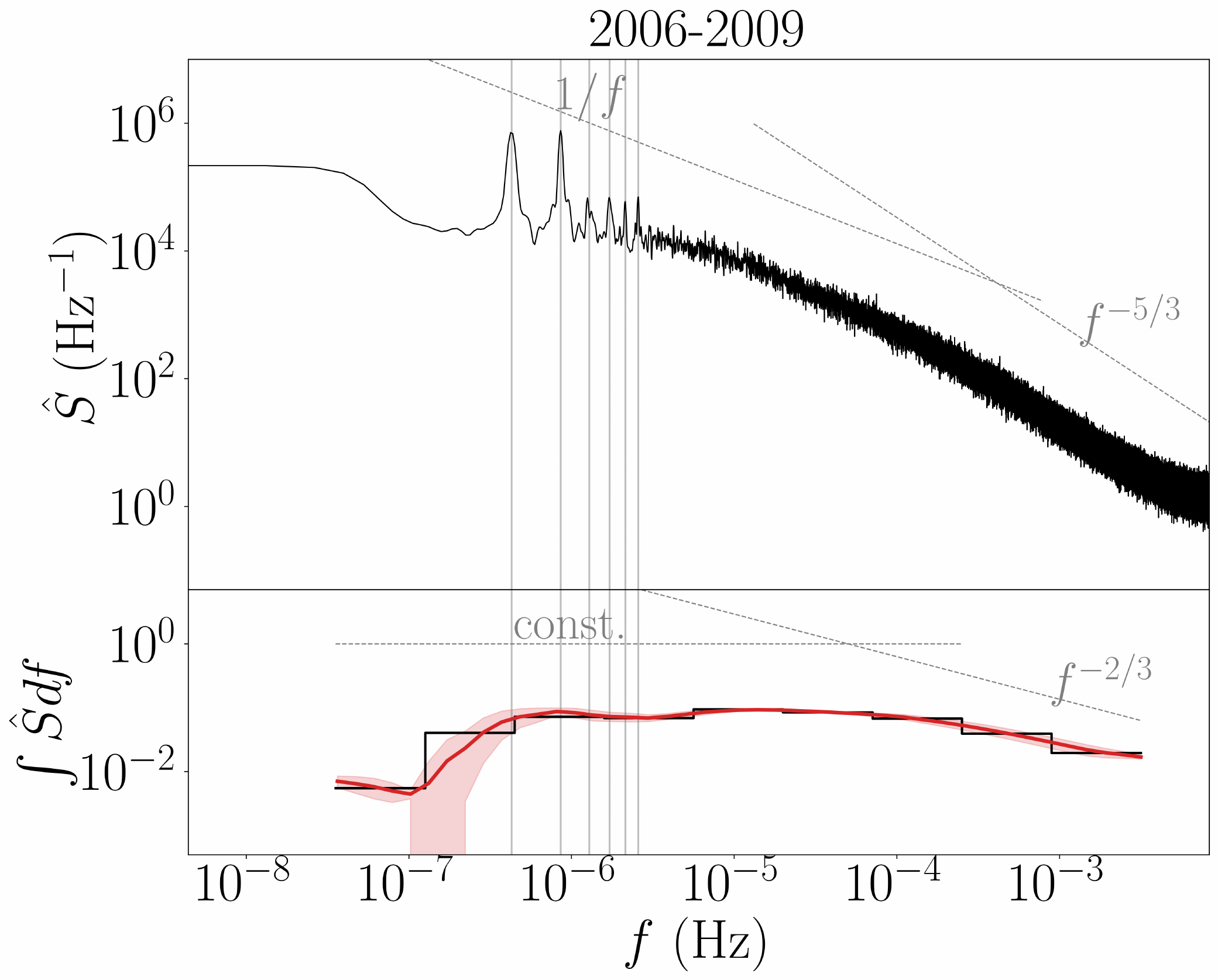}
    \caption{Top panels in each plot: Power spectral densities of magnetic field components (left) and normalized trace spectra (right) for full 12-year dataset, as well as for slow wind, fast wind, solar maximum, and solar minimum subsets. Vertical gray lines denote 27-day solar rotation frequency and its superharmonics. Dashed reference lines indicate $1/f$ and $f^{-5/3}$ power-laws. Bottom panels in each plot: Solid lines show integrated spectra estimated using sliding logarithmically spaced frequency bins, with shaded regions in matching colors representing uncertainties. Black piece-wise constant horizontal bars show integrated spectra estimated using fixed frequency bins. Dashed reference lines indicate constant and $f^{-2/3}$ power-laws, corresponding to $1/f$ and $f^{-5/3}$ scalings in spectral density, respectively.}
\label{fig:bspec}
\end{figure*}

The panels of Fig.~\ref{fig:bspec} illustrate 
power spectral densities of magnetic field components (left column) and the normalized trace spectrum (right column) computed from the autocorrelations. 
The rows show spectra from each of the five analyzed wind categories: (1) the full 12-year dataset, (2) slow wind, (3) fast wind, (4) the solar 1999-2002 maximum, and (5) the 2006-2009 solar minimum. 
Each panel is divided into the corresponding spectral
density itself (top half), and the 
associated 
coarse-grained (integrated in frequency bins) 
spectra (bottom half). Compensated trace spectra $f\hat{S}$, plotted in both log-log and linear-log formats, are provided in the SI Appendix.

For the radial and tangential spectra (in the upper part of each panel), and correspondingly in the trace spectrum, a pronounced peak in power exists at the frequency associated with the $\sim 27$ day solar rotation period, with additionally peaks at its superharmonics. Notably, in both the slow wind and the solar minimum cases, the spectral power at the first superharmonic at 13.5 days exceeds that at 27 days, consistent with the secondary group of maxima at 13.5-day periodicity readily observed in their respective correlation functions. Additionally, near the 27-day peak, the slow wind spectrum contains a nearby peak at around 30 days, a possible source of the beat-like interference. Meanwhile, another possible source is the intensified 27-day peak broadening seen in the solar maximum. In both these cases, 
beat-like interference due to 
closely spaced frequencies may underlie the slight periodicity drift in the corresponding correlations (see top panels of Figs.~\ref{fig:bcorr_minmax} and~\ref{fig:bcorr_slowfast}). This drift emerges 
earlier in lag (at smaller yearly timescale) in the slow wind and in the solar maximum cases 
relative to the others.

To examine spectral scaling across different scales and to probe the extent of the $1/f$ range, we estimate the integrated spectra from the spectral densities using the method outlined in Section~\ref{sec:data}. The integrated spectra are plotted in the bottom panels of Fig.~\ref{fig:bspec}. The inertial range turbulence spectrum with a $f^{-5/3}$ power-law corresponds to a $f^{-2/3}$ scaling in the integrated spectrum, whereas the $1/f$ region appears flat, reflecting its characteristic property of equal power per octave~\citep[][]{Machlup81}.

Across all five cases analyzed - namely the full 12-year dataset, slow wind, fast wind, solar minimum, and solar maximum - the radial and tangential magnetic field exhibit generally similar spectral behavior over the full range of frequencies studied (approximately $\unit[10^{-8}]{Hz}$ to $\unit[3 \times 10^{-3}]{Hz}$, corresponding to timescales from around $3\unit[300]{seconds}$ to $\unit[3]{years}$). One exception is in the fast wind between $\unit[1.3 \times 10^{-6}]{Hz}$ (the second superharmonic of solar rotation) and $\unit[10^{-5}]{Hz}$, where a slight deviation occurs. In general, the spectral indices for the radial and tangential components hover around $-1$ or marginally steeper immediately above the solar rotation frequency for at least a decade, before transitioning into the $f^{-5/3}$ inertial range. The normal component, by contrast, tends to remain flat - indicative of uncorrelated signals - before directly transitioning into the $f^{-5/3}$ regime, bypassing a clearly defined $1/f$ range. Therefore, the $1/f$ behavior observed in the trace spectrum for all five cases likely arises from the collective contribution of all three magnetic field components. 

\begin{figure}[t]
\centering
    \includegraphics[angle=0,width=\columnwidth]{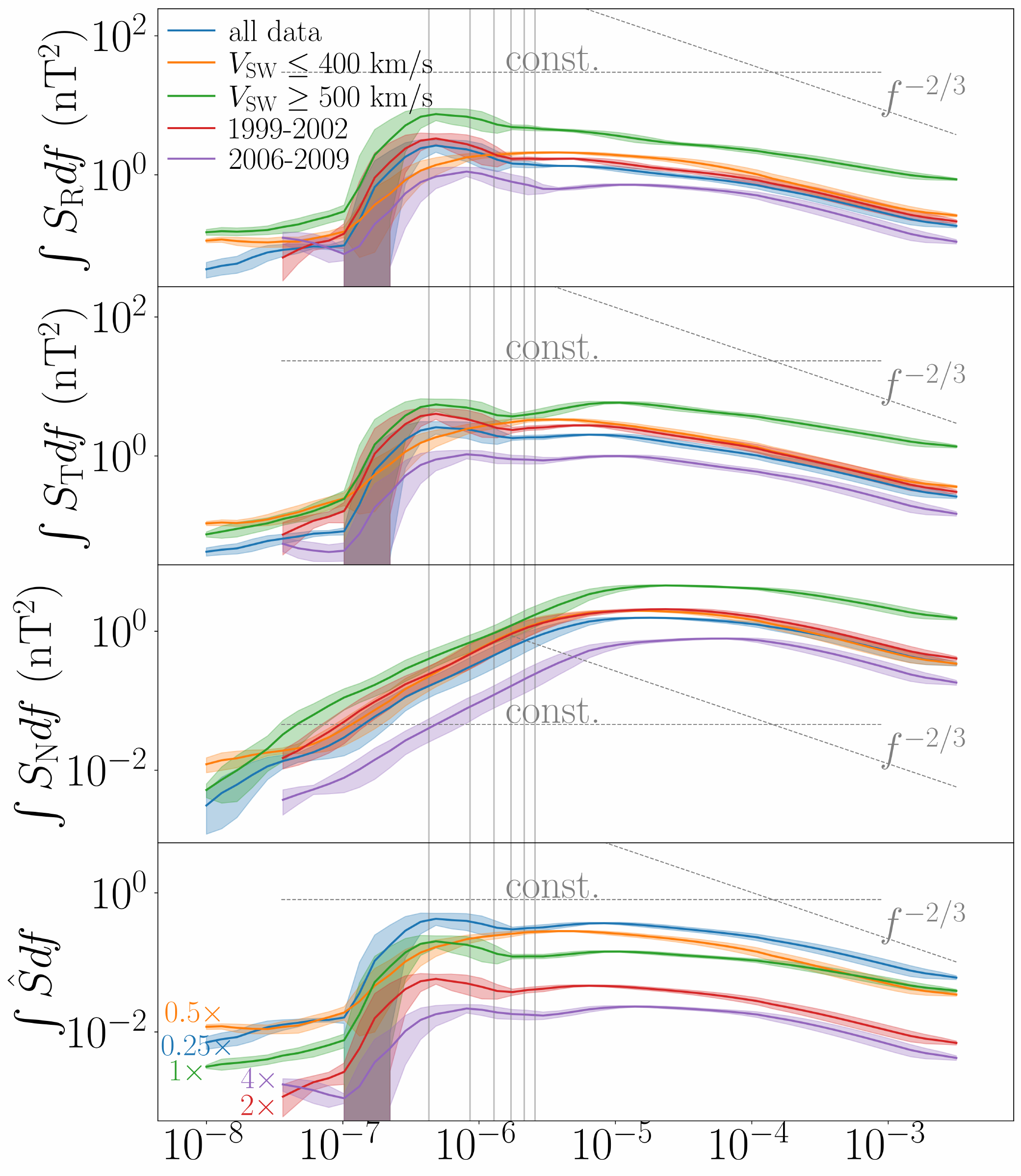}
    \caption{Integrated spectra for radial, tangential, normal magnetic field components and integrated normalized trace spectrum (from top to bottom), estimated using sliding logarithmically spaced frequency bins, with shaded regions in matching colors representing uncertainties. Dashed reference lines indicated constant and $f^{-2/3}$ power-laws, corresponding to $1/f$ and $f^{-5/3}$ scalings in spectral density, respectively. Spectra in bottom panel are multiplied by indicated factor for visualization.}
\label{fig:bintS}
\end{figure}

In Fig.~\ref{fig:bintS}, we compare the component-wise integrated spectra and their normalized traces across all five analyzed cases. In all trace spectra, the $1/f$ regime extends well below the previously observed lower bound of around $\unit[10^{-6}]{Hz}$~\citep[see, e.g., ][]{Matthaeus86}, and continues through the frequency range populated by solar rotation superharmonics. In the cases of the entire 12-year dataset as well as during the solar minimum, the $1/f$ behavior persists down to the fundamental solar rotation frequency. This finding challenges the earlier conjecture that large-scale periodic structures associated with solar rotation and its superharmonics disrupt $1/f$ signatures at comparatively smaller timescales~\citep[see, e.g., ][]{Matthaeus86, Dorseth24}.

Toward the high-frequency end, the transition from the $1/f$ to the Kolmogorov inertial range occurs at different frequencies depending on the solar wind category. For the solar maximum and slow wind cases, the transition typically occurs near $\unit[10^{-5}]{Hz}$. However, for the solar minimum and fast wind cases, the transition shifts to approximately a decade higher in frequency. The variation of the transition frequency with wind speed is consistent with the findings of \cite{Tu95, Bruno13, Bruno19}, that in the slow wind that transits slower toward Earth compared to the fast wind, turbulence structures have more time to develop. This produces in slow wind a more expansive Kolmogorov range. Thus the $1/f$ range is encountered at a relatively lower range of frequencies in slow wind. 

As seen in the component-wise spectra, the power in magnetic field fluctuation is consistently greater in fast wind intervals compared to slow wind, and greater during solar maximum relative to minimum. Notably, the slow wind shows suppressed power across the frequency range corresponding to the solar rotation and its superharmonics, and has excess power at subsequently higher frequencies before transitioning to a $f^{-5/3}$ inertial range. In contrast, the fast wind inertial range spectrum for all three magnetic field components maintains a shallower spectral scaling at relatively higher frequencies.

\subsection{Density correlation and spectrum}

\begin{figure*}
\centering
    \includegraphics[angle=0,width=0.85\columnwidth]{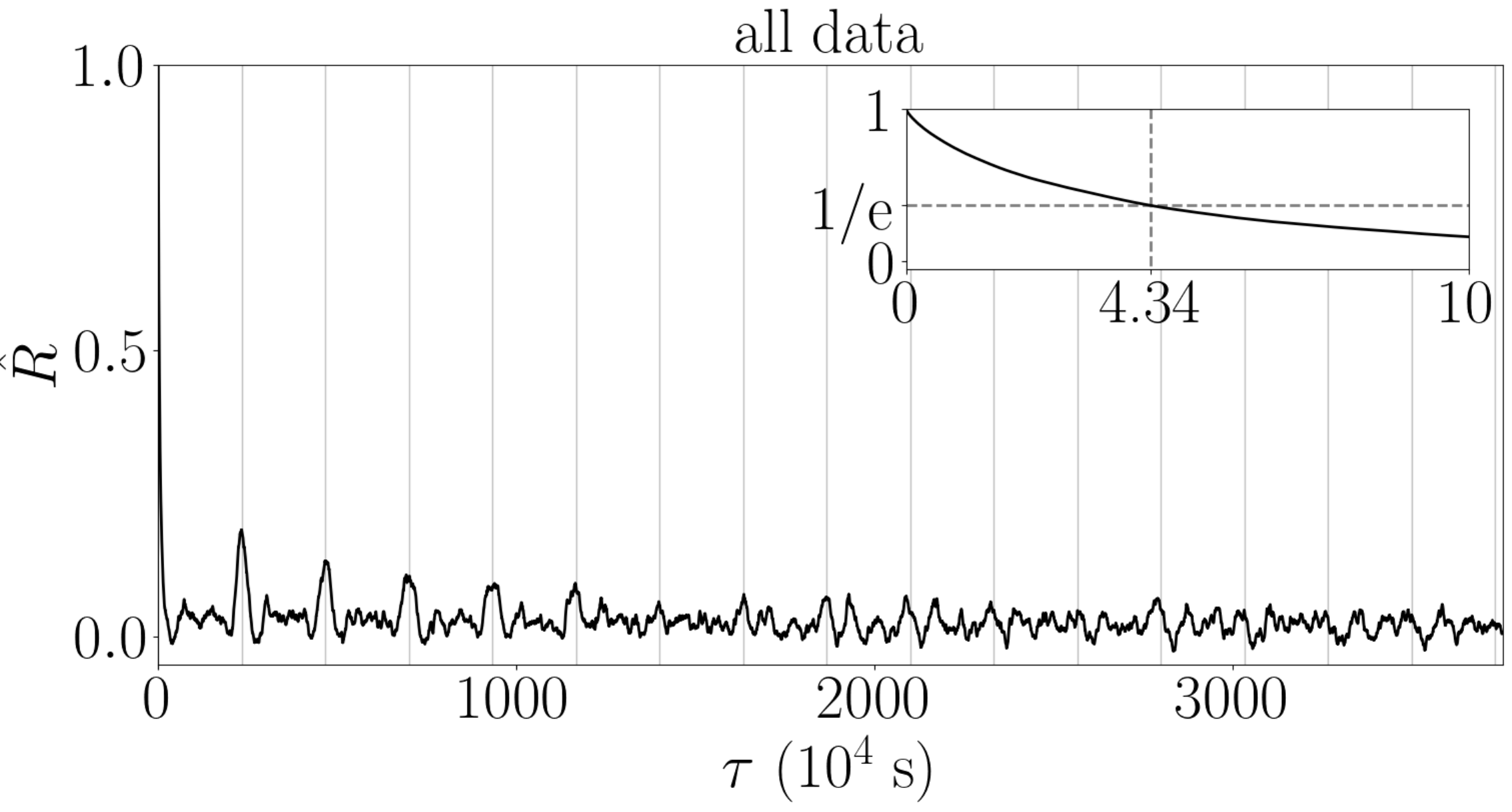}
    \includegraphics[angle=0,width=0.85\columnwidth]{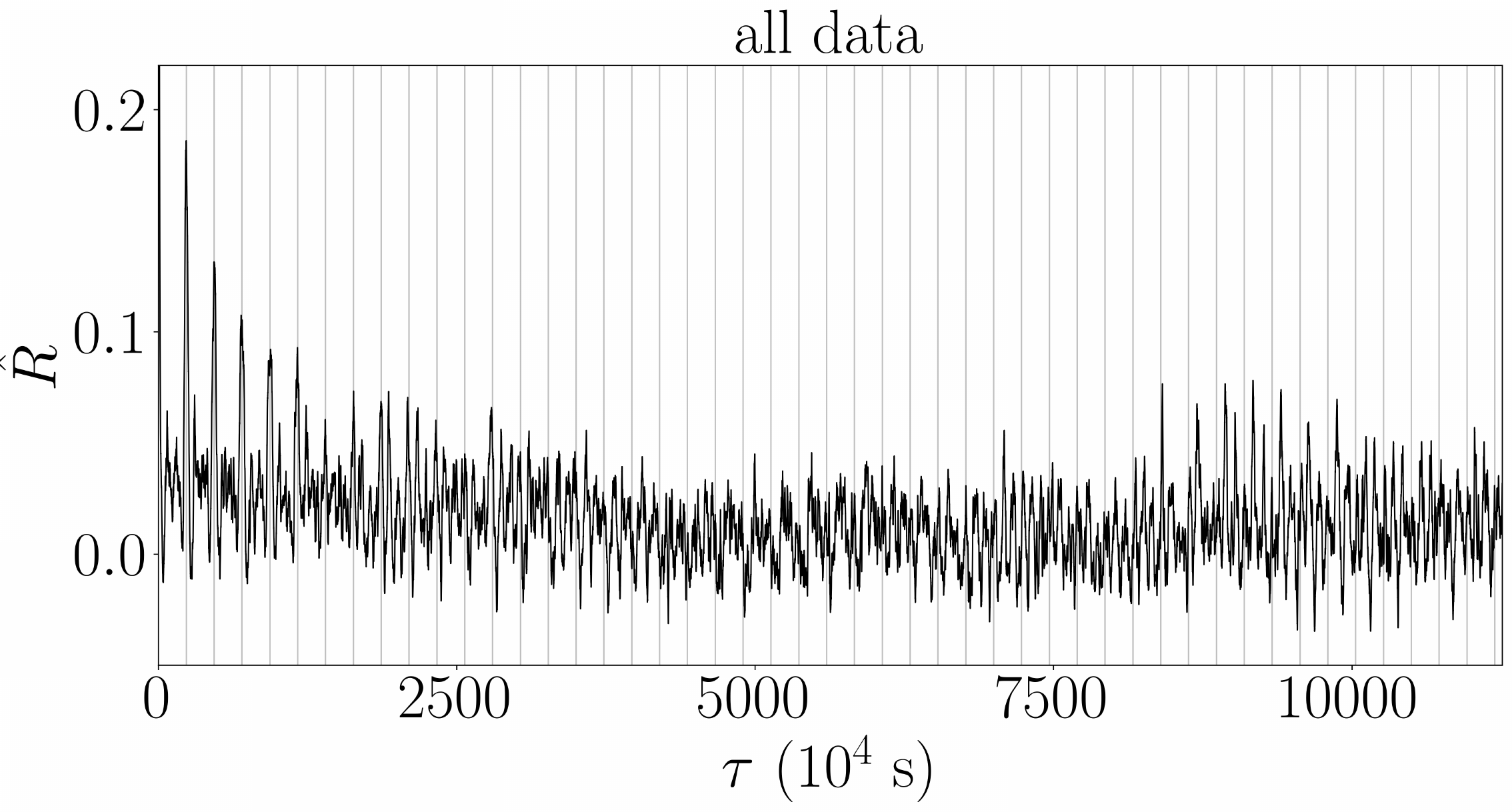}
    \caption{Normalized density autocorrelation of full 12-year data shown up to 1.2 year (left) and 4 years (right) temporal lag. Vertical gray lines mark integer multiples of 27 days. Inset shows magnified view of correlation function up to around its correlation ($e$-folding) scale, which is indicated by vertical dashed line. Vertical scale in right panel is adjusted for clarity and zero-lag correlation maxima is not shown.}
\label{fig:ncorr}
\end{figure*}

\begin{figure}
\centering
    \includegraphics[angle=0,width=0.85\columnwidth]{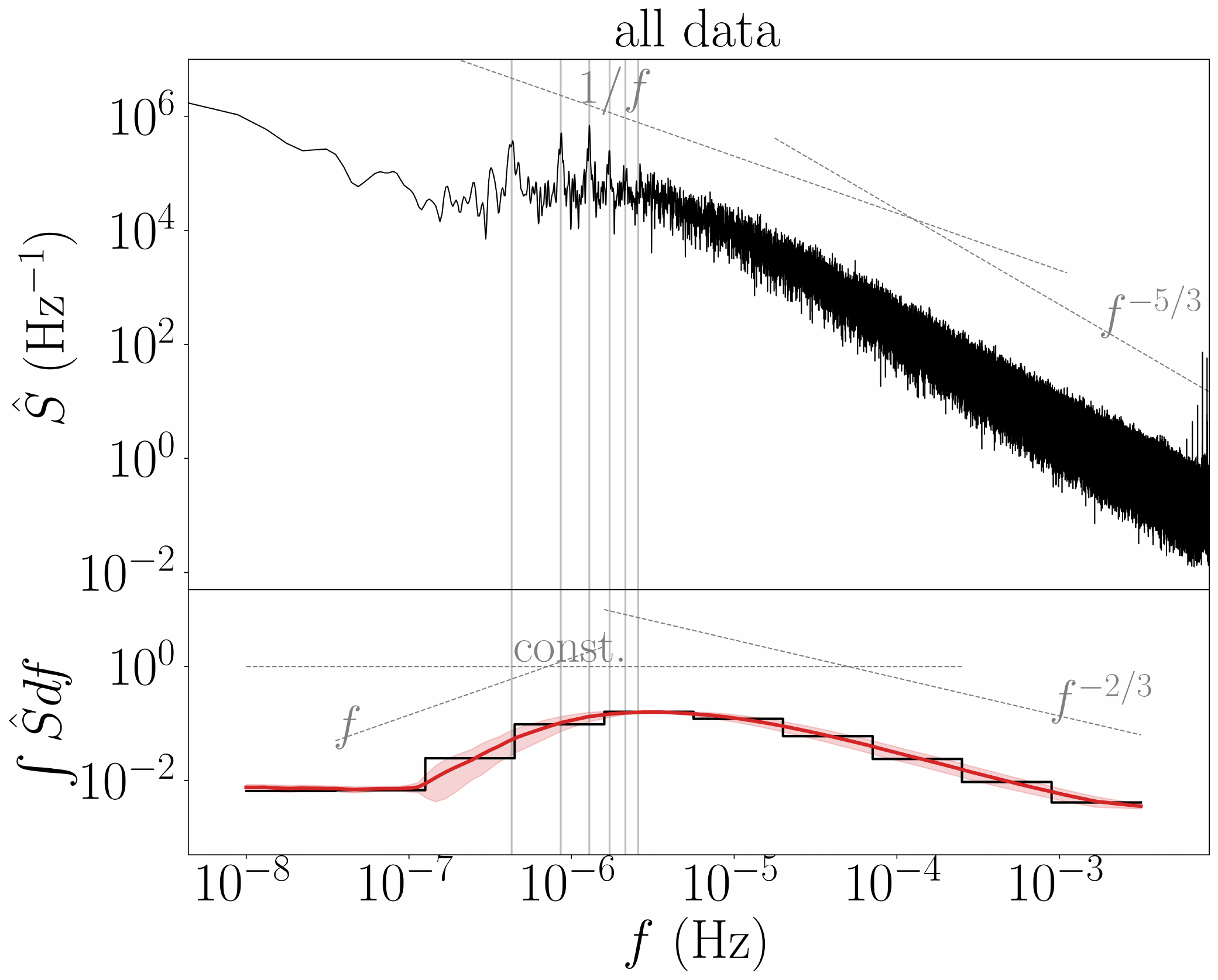}
    \caption{Top panel: Normalized power spectral density of full 12-year density data. Vertical gray lines denote 27-day solar rotation frequency and its superharmonics. Dashed reference lines indicate $1/f$ and $f^{-5/3}$ power-laws. Bottom panel: Solid red line shows integrated spectra estimated using sliding logarithmically spaced frequency bins, with shaded regions in matching colors representing uncertainties. Black piece-wise constant horizontal bars show integrated spectra estimated using fixed frequency bins. Dashed reference lines indicate $f$, constant, and $f^{-2/3}$ power-laws, corresponding to flat, $1/f$, and $f^{-5/3}$ scalings in spectral density, respectively.}
\label{fig:nspec}
\end{figure}

To complement the magnetic field analysis, we present the autocorrelation and power spectrum of solar wind density computed from the full 12-year dataset, shown in Figs.~\ref{fig:ncorr} and~\ref{fig:nspec}, respectively. The turbulence correlation time (here defined as the $e$-folding time) for density is approximately 12 hours, about 1.6 times longer than that of the magnetic field. However, the exact value of this correlation time should be interpreted with caution, since the correlation time is known to be sensitive to the duration of the data interval~\citep[][]{Isaacs15}. When extending the correlation computation to time lags of up to 3.6 years, a beat-like phenomenon emerges, as seen more prominently in the magnetic field.

The density data displays strong correlation at 27-day lag and its first four to five integer multiples. In addition, correlations recurring at sub-27-day lags suggest the presence of structure appearing roughly three times per solar rotation. This is corroborated by the power spectral density, where the second superharmonic ($\sim 7$ days) contains the highest power among all spectral peaks. This behavior contrasts with that of the magnetic field, whose dominant spectral peaks are at the fundamental (27-day) and first superharmonic (13.5-day) frequencies, with comparatively diminished power at higher-order harmonics.

The density spectrum appears uncorrelated across the frequency range encompassing the solar rotation frequency and its harmonics. It then gradually transitions into the $f^{-5/3}$ scaling characteristic of inertial range turbulence, without going through a distinct $1/f$ regime. However, below the solar rotation frequency, a region consistent with $1/f$ scaling develops below approximately $\unit[10^{-7}]{Hz}$, spanning at least a decade in frequency. Resolving the full extent of this ultra-low-frequency $1/f$ band will require analysis of longer time series than employed in the current study. 

\section{Conclusion and Discussion}
\label{sec:conclusion}

Our purpose has been to study 
in some detail 
the behavior of long-timescale, very-low-frequency interplanetary 
turbulence. 
A particular emphasis has been the connection
between the fluctuation properties from the 
nominal correlation scale ($\sim$ one hour)
to the nominal solar rotation scale
($\sim$ 27 days).
To this end, we compute two-time correlation functions 
and spectra of the interplanetary magnetic field and proton density using a 12-year sample of ACE data.
The focus has not been on structure in wavenumber, but rather structure in frequency, 
so we have not invoked the Taylor frozen-in hypothesis 
generally used 
to justify conversion to wavenumber~\citep{Jokipii73}. 
Working in frequency space, we 
do not attempt refined commentary on the 
spatial structure of the inertial range, usually expected to have a wavenumber dependence close to 
a Kolmogorov-like spectral index of $-5/3$.
Nevertheless, we can recognize the inertial range 
form as near $f^{-5/3}$, 
in essentially all wind categories but the fast solar wind (see below).
We also note that toward low frequencies at around $10^{-5}$ to $\unit[10^{-4}]{Hz}$, the 
spectral form gradually 
transitions to the $1/f$ regime.
This extended transition region is consistent with the use of a long data record, 
along with the established large-variance log-normal distribution of interplanetary magnetic field correlation times~\citep{Ruiz14, Wang24_1overf} and the nature of the superposition principle~\citep{Machlup81, Montroll82}
that provides a generic pathway to the generation of $1/f$ signal. 

In the fast solar wind ($V_\mathrm{SW} \geq \unit[500]{km/s}$), the inertial range spectral index -- typically observed at frequencies above around $\unit[3 \times 10^{-4}]{Hz}$ (if considering the lower bound of the 1 au correlation time distribution at around 1 hour) -- deviates from $-5/3$ and flattens. This trend is evident across all three magnetic field components. Evidence of spectral flattening in fast wind has been presented by \cite{Borovsky12} employing ACE intervals categorized by wind speed and wind origin (e.g. coronal-hole, non-coronal-hole, and ejecta based on ionized oxygen ratio). Magnetic field spectral index is systematically less negative in fast wind and in wind of coronal-hole origin. \cite{Borovsky12} also identifies intervals of sustained shallow spectra reaching a spectral index of -1.34, which tend to have substantially enhanced velocity and magnetic field fluctuations, Alfv\'en ratios, and outward-to-inward Els\"asser ratios. \cite{Teodorescu15} using measurements at 0.72 au also find spectral flattening in fast wind, and attributes the observed slope tending toward $-3/2$ to strong Alfv\'enicity of the flow and limited turbulence development due to fast wind's rapid transit. More recently, \cite{Sioulas23} reports spectral indices near $-3/2$ independent of plasma speed closer to the Sun using Parker Solar Probe data.

At long time lags, the 
most conspicuous feature of the
two-time magnetic correlation
is the recurrence of correlation 
in the tangential $\mathrm{T}$
and radial $\mathrm{R}$
components at integer multiples of around 27 days. 
The normal $\mathrm{N}$ component shows no such 
quasi-periodic behavior.  
While a perfect recurrence of correlation at multiples of 27 days correspond to a single frequency in the power spectrum, the observed modulation of correlation near 27 days with a year-long envelope may be accounted for by the presence of nearby frequencies. 
One may readily demonstrate (not shown here) this "beat"-like phenomenon with a simple test using synthetic data. One form of data behavior is the wave packet configuration proposed by \cite{Fenimore78}, where packets of signals with an exact 27-day periodicity are phase shifted by various amounts  that results in multiple periodicities around 27 days. This mimics coronal structures that generate recurrence of fluctuations per solar rotation, that appears at varying solar longitudes.

Besides the usually dominant 27-day pattern, the correlation also shows notable recurrence near a 13.5-day cadence, with its prominence varying with wind speed and solar cycle phase. 
In particular, the 13.5-day harmonic is most apparent during solar minimum, exhibiting greater power than the 27-day signal. In contrast, the solar maximum correlation lacks clear secondary periodicity, but gradually drifts out of phase with the 27-day cadence after around 5 solar rotations. The fast wind correlation recurrence at 27 days persists up to around one year in time lag, after which a 13.5-day structure emerges. For slow wind, both 13.5-day and higher-order fluctuations appear after around 4 solar rotations, interfering with and obscuring the ``beat''-like modulation observed in the fast wind and full 12-year analyses. Overall, we identify two distinct features associated with coronal-origin periodicities: (1) subdominant correlation maxima related to harmonics of the 27-day periodicity, shaping solar minimum correlation; and (2) drifting correlation maxima and envelope modulation in their amplitudes arising from tightly clustered frequencies, shaping solar maximum correlation. Slow and fast wind correlations seem to assimilate both mechanisms.

Several studies have addressed the observed periodicities in the solar wind.
A recent example is \cite{Dorseth24}, which studies the
slow wind magnetic field from the {\it Wind} spacecraft. In addition to a clear 27-day periodicity in their magnetic field correlation trace, they find a $1/f$ spectral regime consistent with our results. However, they suggest that the lower frequency boundary the $1/f$ range is at around $\unit[4\times 10^{-6}]{Hz}$, truncated by solar rotation harmonics. 

Despite broad agreement on the existence of these periodicities, there lacks a consensus as to whether harmonics such as at 13.5 days are features of solar dynamics or are associated with 
large interplanetary structures 
such as the heliospheric current 
sheet. 
The latter explanation would seem to be consistent with the presence of a wavy current sheet~\citep[i.e., the ``ballerina skirt'';][]{Burlaga95} at solar minimum, with a two-sector structure.
Then any near-Earth spacecraft would
detect two current sheet crossings and the associated magnetic polarity change roughly every 13.5 days. 
Therefore, while this explanation depends on interplanetary structures, these structures at large scales must reflect processes within the Sun, or at least in the low solar atmosphere. From this perspective, the 13.5-day 
harmonic must be fundamentally linked to solar dynamics and their kinematic
extensions into the interplanetary space, 
and is not due,
and least not in a principal way, 
to interplanetary dynamics.

Although correlation and spectrum offer equivalent content, the spectrum provides additional insight, particularly into the $1/f$ behavior and its relationship to the strengths of the 27-day harmonics. 
Perhaps the most remarkable and
unexpected result in the present study
is that when power is integrated in frequency 
bins ranging over the 27-day spectral peak and its harmonics, its distribution blends smoothly into the neighboring $1/f$ continuum.
We are thus led to the question: can the point spectral features at 27 days and related periods be consequences of the same processes that produce the broadband $1/f$ signal?
Our results suggest that these harmonics in the low-frequency range
have a closer connection to the 
$1/f$ signal than previously thought.

A compelling hypothesis is that the observed $1/f$ noise (in the time domain) may originate within the solar dynamo (which admits specific spatial scales).  $1/f$-type fluctuations are often found in dynamo simulations and laboratory experiments, in analogy with terrestrial and planetary dynamos~\citep[see, e.g.,][and references therein]{Dmitruk14}. While explicit evidence in solar dynamo is less expected due to its 11-year quasi-periodicity, notable examples exist, such as \cite{Polygiannakis03} that report a $1/f$ signal in long-term sunspot records, and \cite{Kostuchenko98} that find $1/f$ in solar mean magnetic field, solar irradiance, and solar neutrino capture rate. 

One possible explanation is that inverse cascade processes occur in solar dynamo, potentially supported by the presence of pseudo-invariants~\citep[see][and references therein]{Dmitruk07}. Numerical studies show that systems undergoing inverse cascade often develop $1/f$ spectra, through competition with direct cascades~\citep{Consolini15} or through the generation of timescales much longer than the characteristic bandlimited scales (e.g., the eddy turnover time of the largest structure) in a self-organized manner commonly associated with $1/f$ noise~\citep{Ponty04, Dmitruk11}.

The connection of the solar inverse cascade process to signals measured in the time domain is not entirely well understood, but may be connected to self-organized criticality~\citep[SOC;][]{Bak87, Bak88} -- a framework describing how systems under small perturbation naturally evolve toward dynamical cluster of minimally stable states with no characteristic spatial and temporal scales. The emergence of a $1/f$ signal in such a framework, possibly including in the dynamo, may be related to interaction across scales~\citep{Hwa92}, anomalous relaxation processes~\citep{Weron91}, as well as finite size effects~\citep{Korzeniowska23}.

In contrast to the magnetic field that exhibits correlation near the band of solar rotation harmonics, the plasma density spectrum reveals the behavior of the fundamental and the first two harmonics superimposed on an almost completely uncorrelated background. The density spectrum then transitions directly into a $f^{-5/3}$ inertial range. Around half a decade of the transition region displays a $1/f$-like scaling; however, this may simply reflect the gradual nature of the decade-long transition as the result of long-duration dataset employed. It is worth noting that $1/f$ spectrum in the interplanetary density has been reported by \cite{Matthaeus07} near $\unit[10^{-4}]{Hz}$, but only at mid- to high-solar latitudes dominated by fast solar wind; such scaling is absent in ecliptic plane measurements. The model proposed by \cite{Magyar22} for generating $1/f$ Alfv\'en fluctuations relies on a transversely inhomogeneous density background possibly rooted in the solar corona. The relationship between density and magnetic field $1/f$ signatures remains incompletely understood. At a basic level, our analysis suggests that the mechanism responsible for producing the well-established $1/f$ behavior in the magnetic field has little impact on the density field near the ecliptic, at least on the year-long timescales probed by our data record. 

We have presented an in-depth study of a
long data record of magnetic field 
from the ACE spacecraft, 
analyzing both correlation functions and spectra. 
Use of the mean-lagged product method permits analysis conditioned on 
phase of solar cycle and wind speed. 
What emerges is a surprisingly broad region of $1/f$, if one accepts our suggestion that 
the power at the solar rotation harmonics, when coarse grained, is consistent with an extension of the $1/f$ signal down to a few times $\unit[10^{-7}]{Hz}$.
This contrasts with previous interpretation of a low-frequency limit of a few times $\unit[10^{-6}]{Hz}$, based on the idea that the rotation harmonics are of different origin, 
and {\it interfere} with the $1/f$ rather than contributing to it.
This viewpoint, of course, remains to be 
further studied, and may potentially be supported by research on the 
dynamical processes within the Sun or in the lower corona that contribute to the 27-day signal and its harmonics.

The high-frequency end of the $1/f$ range exhibits 
a gentle rollover to the anticipated Kolmogorov-like scaling associated with 
the classical turbulence inertial range. The gradual nature of the rollover is 
connected in an essential way to the 
distribution of {\it local} 
correlation scales, which may vary greatly~\citep{Ruiz14, Isaacs15}.
By the {\it superposition principle},
it is the log-normality of the correlation scales that gives rise to the $1/f$ scaling within the frequency range over which this correlation distribution is scale-invariant.
In the upper-frequency range when the scale-invariance drastically weakens, the gradually transition to the classical turbulence range sets in. 
This gentle transition in the entire ensemble can become much sharper in 
analysis of small data subsets. 

On the other hand, previous work~\citep{Isaacs15} has shown that long data records, such as those used in the present analysis, tend to exhibit long e-folding times in their inertial range autocorrelations. This behavior likely reflects that the underlying decay is not purely exponential, but better described by a stretched exponential form, consistent with the presence of a broad range of correlation scales. Furthermore, within the framework of self-organized criticality~\citep{Bak87, Hwa92}, interactions among these timescales possibly associated with dynamics in the solar dynamo may produce even longer scales (reminiscent of inverse cascade processes), and thereby sustain an extended regime of $1/f$ spectrum.

Recently, there have been a number of studies that seek to explain interplanetary $1/f$ observations
as a local phenomenon~\citep{Huang23, Davis23} possibly involving expansion effects~\citep{Velli89, Verdini12}.
While the role of local dynamics cannot be ruled out at relatively higher frequencies (above a few times $\unit[10^{-5}]{Hz}$), such mechanisms 
clearly fail to explain the signals below around $\unit[10^{-6}]{Hz}$ due to limitations posed by the {\it range of influence}~\citep{Zhou90,Chhiber2018thesis,Wang24_1overf}. 
If such local processes are operative, the problem remains 
as to how to smoothly connect the associated $1/f$ spectra to the 
lower-frequency part we have studied 
here. 
Further studies including 
simulations, 
multispacecraft 
observations such as by HelioSwarm~\citep{Klein23}, and 
remote-sensing observations such as by the new PUNCH mission~\citep{Deforest22} may help to clarify these issues. 



The velocity and density data were downloaded from \url{https://spdf.gsfc.nasa.gov/pub/data/ace/swepam/level2_hdf/ions_64sec}. The magnetic field data were downloaded from \url{https://spdf.gsfc.nasa.gov/pub/data/ace/mag/level_2_cdaweb/mfi_h3/}. This research is partially supported by the NASA LWS grants 80NSSC22K1020, by the NASA IMAP project at UD under subcontract SUB0000317 from Princeton University, by the NASA/SWRI PUNCH subcontract N99054DS, and by National Science Foundation grant AGS-2108834.


\end{CJK*}

\end{document}